\documentclass[aps,prd,twocolumn,superscriptaddress,nofootinbib,floatfix]{revtex4-2}

\usepackage[T1]{fontenc}
\usepackage[utf8]{inputenc}
\usepackage{lmodern}
\usepackage[reqno]{amsmath}
\usepackage{amssymb,bm,mathtools}
\usepackage{graphicx}
\usepackage{booktabs}

\usepackage{multirow}
\usepackage{xcolor}
\usepackage{tikz}
\usetikzlibrary{arrows.meta,positioning,calc,matrix}
\usepackage{pgfplots}
\pgfplotsset{compat=1.18}
\usepackage{hyperref}
\usepackage{silence}
\hypersetup{colorlinks=true,linkcolor=blue,citecolor=blue,urlcolor=blue,
  pdftitle={Unified Flavor: Lattice Quantization, Chain Locality, and a Dynamical Origin of Hierarchical Yukawas},
  pdfauthor={Vernon Barger}}
\usepackage{bookmark}

\allowdisplaybreaks

\newcommand{\e}{\epsilon}
\newcommand{\Bpar}{B}
\newcommand{\UF}{Unified Flavor}
\newcommand{\Z}{\mathbb{Z}}
\newcommand{\diag}{\mathrm{diag}}

\newcommand{\Ye}{Y_e}
\newcommand{\Ue}{U_{eL}}
\newcommand{\Unu}{U_\nu}

\begin{document}

\title{Unified Flavor:\\Lattice Quantization, Chain Locality,\\and a Dynamical Origin of Hierarchical Yukawas}

\author{Vernon Barger}
\affiliation{Department of Physics, University of Wisconsin--Madison,
Madison, Wisconsin 53706, USA}

\date{March 11, 2026}

\begin{abstract}
We present \UF{} (UF), a framework that synthesizes
the $\Bpar$-lattice flavor hierarchy with a dynamical
realization based on TeV-scale vectorlike fermion (VLF) chains.
Hierarchical Yukawa couplings arise from discrete
ninths-quantized lattice exponents enforced by a single
flavon $\Phi$ with $\e\equiv\langle\Phi\rangle/\Lambda=1/\Bpar$,
$\Bpar=75/14$.
Effective Yukawa entries are generated as algebraic path sums
along nearest-neighbor chains of vectorlike quarks (VLQs),
factorizing into entry, chain-propagation, and exit amplitudes
controlled by the discrete gauge charges.
A multi-messenger structure---in which each Yukawa entry
receives coherent contributions from several chain
configurations---generates $\mathcal{O}(1)$ complex
coefficients whose phases are the physical origin of
CP violation.
We derive a general chain-inversion theorem, perform systematic
perturbative diagonalization of both up- and down-type
Yukawa textures, and show that the
Cabibbo--Kobayashi--Maskawa (CKM) mixing
hierarchy and CP-phase structure emerge naturally from the
lattice exponent algebra and multi-messenger interference.
All six quark masses are reproduced with
$\mathcal{O}(1)$ coefficients that are essentially unity.
The chain locality simultaneously suppresses dangerous
flavor-changing neutral currents (FCNCs) and satisfies
electroweak precision constraints, while requiring
VLQs with masses in the multi-TeV range accessible at
the High-Luminosity Large Hadron Collider (HL-LHC).
The same discrete gauge symmetry that enforces the lattice
structure also protects the Peccei--Quinn axion quality,
unifying flavor, CP violation, and the strong CP problem.
The framework extends to the lepton sector, reproducing
charged-lepton mass hierarchies, the normal-ordered neutrino
spectrum, and PMNS mixing with a predictive two-branch
octant--$\delta$ correlation testable in upcoming experiments.
\end{abstract}


\maketitle

\section{Introduction}
\label{sec:intro}

\bigskip
\noindent
\textit{The core of this Froggatt--Nielsen (FN) flavor phenomenology is
quadrilateral combinations of fermion masses evaluated
at scale $M_Z$ that are found to be quantized as power
laws $\Bpar^n$ on a lattice with a universal constant
$\Bpar=75/14$ and integer quantum numbers $n$.
From these integers $n$, the masses and the weak mixing
elements are inferred to be power laws $(1/\Bpar)^p$
where $\e=1/\Bpar$ is the small FN parameter and the
$p$-values are individually predicted as rational
fractions that are determined by the $n$-values.}
\bigskip

Understanding the origin of fermion mass hierarchies and
flavor mixing remains a central challenge in particle physics.
The Froggatt--Nielsen (FN) mechanism~\cite{Froggatt:1978nt,Leurer1992,Leurer1993}
generates hierarchical Yukawa couplings through powers of a
small expansion parameter $\e=\langle\Phi\rangle/\Lambda$
set by the vacuum expectation value (VEV) of a single flavon
field $\Phi$.
The physical picture is as follows.
Each Standard Model fermion carries a generation-dependent
$\mathrm{U}(1)_{\rm FN}$ charge (or, more generally, a discrete
$Z_N$ charge).
A Yukawa coupling $\bar f_L\, f_R\, H$ is forbidden at tree level
by the charge mismatch between the left- and right-handed
fermions; it is generated only through higher-dimension operators
containing $n$ powers of $\Phi/\Lambda$, where $n$ is fixed by
charge selection rules.
After the flavon develops its VEV the effective Yukawa coupling
scales as $y_{ij}\sim\e^{n_{ij}}$, so that the observed
mass hierarchies---spanning five orders of magnitude in the quark
sector---are converted into modest differences in the integer
(or, once discrete gauge symmetries are admitted, rational)
exponents $n_{ij}$.
In the standard FN framework these exponents are integers
counting flavon insertions along a single messenger line.
The key innovation of the $\Bpar$-lattice program is that
discrete gauge symmetries of the type analyzed by
Ib\'a\~nez and Ross~\cite{IbanezRoss} permit rational exponents
quantized in ninths, enabling a single expansion parameter
$\e=1/\Bpar$ to fit both quark and lepton mass spectra with
no additional free parameters beyond $\mathcal{O}(1)$
coefficients.

It is important to stress that the fractional powers appearing in
the effective field theory (EFT) do not appear in the ultraviolet
Lagrangian itself.
The UV completion is fully renormalizable: one introduces heavy
vectorlike messenger fermions with mass $M$, and the light
Standard Model fields couple to these messengers and to the
flavon $\Phi$ through ordinary Yukawa
interactions~\cite{Froggatt:1978nt,Leurer1992,Leurer1993}.
Integrating out a chain of $n$ messenger fields generates an
effective operator proportional to
$(\langle\Phi\rangle/M)^n$~\cite{Barger2025bfnb}.
In the $\Bpar$-lattice framework the discrete $Z_N$ charge
differences between adjacent chain sites need not be unity;
they can take values such as $1$, $2$, or $4$ (mod~9), so that
each hop contributes a fractional exponent $h_k/9$ while still
requiring an integer number of flavon insertions $h_k$ at
that vertex.
The ``ninths'' exponents are therefore a compact EFT bookkeeping
for short messenger chains---typically three links---whose
individual hops are governed by discrete gauge charge selection
rules~\cite{IbanezRoss}.
Sections~\ref{sec:field-content} and~\ref{sec:chain-theorem}
below make this correspondence explicit.

It is worth emphasizing that the numerator of a ninths
exponent is \emph{not} the chain length: it is the sum of
discrete charge differences along the chain.
If one simply assigned the flavon a charge of $1/9$, then
integrating out $n$ messengers would give $\e^{n/9}$ with
$n$ equal to the number of links.
That would require 17~messengers to produce
$|V_{cb}|\sim\e^{17/9}$, or 30 for
$|V_{ub}|\sim\e^{10/3}$---hardly economical.
In the $\Bpar$-lattice framework the flavon carries charge~1
under $\Z_9$, but the messenger sites carry different $\Z_9$
charges, so the charge gap $h_k$ at each nearest-neighbor
hop need not be unity.
For instance, a three-site chain with hop charges
$(h_1,h_2,h_3)=(1,2,4)\;\mathrm{mod}\;9$ yields a total
exponent $(1+2+4)/9=7/9$ from only three links.
Because the set $\{1,2,4\}$ generates all residues mod~9,
different Yukawa entries---corresponding to different
endpoint dressings on the same short chain---access the
full ninths lattice $\{n/9\}$ without requiring long chains.
The denominator~9 thus reflects the order of the discrete
gauge group, while the numerator encodes the charge
assignment pattern at the chain sites.

A series of recent papers~\cite{Barger2025bfn,Barger2025bfnb,LatticeFlavonQuarkMixing,LeptonLattice}
has demonstrated that a specific quantization of the
FN exponents---the $\Bpar$-lattice, with
$\Bpar=75/14$ and $\e=1/\Bpar\simeq 0.187$---organizes
quark and lepton masses, CKM and Pontecorvo--Maki--Nakagawa--Sakata
(PMNS) mixing, and CP violation into a unified framework.

Independently, theory-space locality and clockwork/chain
constructions have shown that exponential hierarchies can
emerge from nearest-neighbor couplings among vectorlike
fermions~\cite{Giudice:2016yja,Kaplan:2015fuy,ArkaniHamed:2001is}.
In a recent development, Arkani-Hamed, Figueiredo, Hall,
and Manzari (AFHM)~\cite{ArkaniHamed:2026chains} have
proposed a concrete class of TeV-scale chain models in
which locality in the VLQ chains simultaneously generates
hierarchical Yukawa matrices and suppresses flavor- and
CP-violating processes, with a natural extension to
neutrino masses.
Their framework shares the essential mechanism exploited
here---nearest-neighbor VLQ chains producing effective
Yukawa couplings as products of sequential mass mixings---but
does not incorporate the discrete $\Z_{18}$
lattice quantization that underpins the $\Bpar$-lattice
program and its connection to the axion quality problem.
VLQs have also been studied extensively both as mediators
of flavor and as collider
targets~\cite{Aguilar-Saavedra:2013qpa,delAguila:2000rc,Babu:2025qlunif}.
Discrete symmetry approaches to structured Yukawa textures
provide further motivation~\cite{Altarelli:2010gt,King:2013eh}.

Recent systematic scans of the Froggatt--Nielsen
landscape~\cite{Cornella:2025scan,Cornella:2025pheno} have
explored single-messenger models with integer charges,
mapping their predictions onto the space of quark and
lepton observables.
Babu, Chandrasekar, and Tavartkiladze~\cite{Babu:2026axion}
have independently connected gauged flavor symmetries to
the axion quality problem.
These studies reinforce a key
lesson: with a single flavon and integer FN exponents,
the quark mass hierarchies can be accommodated, but the
precision is limited by the coarseness of the integer
lattice.  The ninths-quantized $\Bpar$-lattice opens a
qualitatively different region of model space---one in
which multi-link chains of vectorlike fermions carrying
discrete $\Z_{18}$ charges produce rational
exponents, yielding substantially better fits to the
observed mass and mixing patterns.  Crucially, the same
discrete gauge symmetry that enforces the lattice
structure simultaneously protects the Peccei--Quinn
symmetry against gravitational spoiling, as shown in the
companion paper~\cite{Barger2025bfn}.

\UF{} (UF) unifies these ideas.
A single flavon enforces the discrete $\Bpar$-lattice
quantization of FN exponents, while TeV-scale
nearest-neighbor VLQ chains generate the effective Yukawa
couplings as path sums of allowed hops.
This framework simultaneously:
\begin{enumerate}
\item generates hierarchical quark and lepton masses from a
single parameter $\Bpar$;
\item produces CKM mixing structures consistent with the
$\Bpar$-lattice flavor
program~\cite{Barger2025bfn,LatticeFlavonQuarkMixing};
\item suppresses dangerous flavor-violating operators via
chain locality;
\item requires TeV-scale VLQs accessible at the HL-LHC;
\item protects Peccei--Quinn axion quality through the
same $\Z_{18}$ discrete gauge symmetry that
enforces the lattice exponents, tying the flavor
hierarchy to the strong CP problem.
\end{enumerate}

\noindent
\emph{Relation to earlier work.}
Papers~I--IV of the $\Bpar$-lattice program established the
lattice structure phenomenologically:
Paper~I~\cite{Barger2025bfn} showed that a single parameter
$\Bpar$ organizes quark and lepton masses;
Paper~II~\cite{Barger2025bfnb} introduced the two-over-two
lattice exponent structure from three messenger chains;
Paper~III~\cite{LatticeFlavonQuarkMixing} derived analytic CKM
phase relations from adjacent exponents; and
Paper~IV~\cite{LeptonLattice} extended the framework to lepton
mixing and the PMNS matrix.
The present work provides the dynamical ultraviolet (UV)
realization: the lattice exponent algebra \emph{emerges} from
chain locality, rendering the hierarchy structural rather than
merely parametric.

Table~\ref{tab:roadmap} summarizes the scope of each paper
in the program.

\begin{table*}[htbp]
\centering
\caption{Roadmap of the $\Bpar$-lattice flavor program.}
\label{tab:roadmap}
\setlength{\tabcolsep}{3pt}
\renewcommand{\arraystretch}{1.05}
\begin{tabular}{clc}
\toprule
Paper & Focus & Ref.\\
\midrule
I & Single-$\Bpar$ mass organization & \cite{Barger2025bfn} \\
II & Two-over-two lattice exponents & \cite{Barger2025bfnb} \\
III & CKM mixing \& FX phase & \cite{LatticeFlavonQuarkMixing} \\
IV & PMNS mixing \& octant--$\delta$ & \cite{LeptonLattice} \\
-- & $\Z_{18}$ axion quality & \cite{FlavorInNinths} \\
V & Chain UV completion (this work) & --- \\
\bottomrule
\end{tabular}
\end{table*}

\section{Discrete Lattice Algebra}
\label{sec:lattice-algebra}

\subsection{Symmetry structure}

We impose the flavor symmetry
\begin{equation}
G = Z_N \times Z_2^{\rm (NN)},
\label{eq:G-symmetry}
\end{equation}
where $Z_N$ (with $N=9$ or $18$) enforces the ninths-quantized
suppression of FN operators, and $Z_2^{\rm (NN)}$ forbids
nonlocal (non-nearest-neighbor) chain couplings.
The discrete gauge anomaly cancellation conditions of
Ib\'a\~nez and Ross~\cite{IbanezRoss} are satisfied
generation by generation for $\Z_9$ or $\Z_{18}$.

A single flavon $\Phi$ carries charge
\begin{equation}
q_N(\Phi) = 1,
\end{equation}
and develops a VEV defining the expansion parameter
\begin{equation}
\e \equiv \frac{\langle\Phi\rangle}{\Lambda}
= \frac{1}{\Bpar},\qquad
\Bpar = \frac{75}{14} \simeq 5.357.
\label{eq:eps-def}
\end{equation}

\subsection{Exponent lattice and algebraic closure}

Allowed Yukawa exponents form the additive lattice
\begin{equation}
\mathcal{L} = \left\{\frac{n}{9}\;\Big|\;n\in\mathbb{Z}\right\},
\end{equation}
so that if two operators carry exponents $a/9$ and $b/9$,
their product carries $(a+b)/9$.
Thus $\mathcal{L}$ is closed under composition.
This algebraic closure ensures that the lattice structure
is preserved under all allowed operator products---a property
that neither continuous-$U(1)$ FN models nor generic clockwork
constructions guarantee.

In the $\Bpar$-lattice program, the charged-lepton and quark
sectors use a refined lattice with denominator $18$
(reflecting the bilinear structure of the Yukawa operator),
while the neutrino Weinberg operator uses
denominator~$9$~\cite{Barger2025bfnb}.
The chain mechanism generates both denominators naturally:
a single chain hop contributes $n/9$, while endpoint dressings
involving both left- and right-handed fields contribute
half-integer ninths, i.e.\ elements of $\mathcal{L}_{18}=\{n/18\}$.

\section{Field Content and Charge Assignments}
\label{sec:field-content}

\subsection{Down-type vectorlike chain}

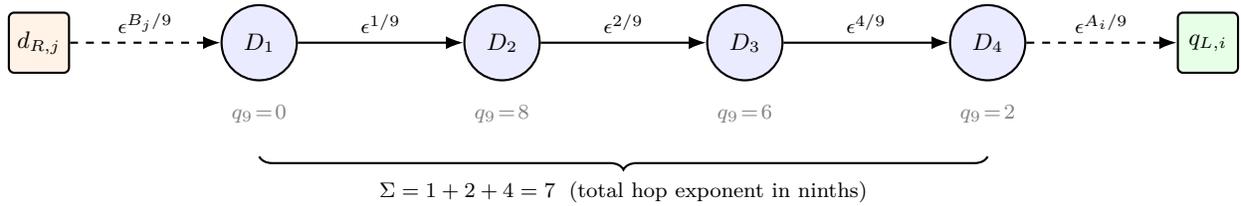
\begin{figure*}[htbp]
\centering
\begin{tikzpicture}[
    node distance=2.2cm,
    site/.style={circle,draw,thick,minimum size=10mm,
                 fill=blue!8,font=\small},
    smL/.style={rectangle,draw,thick,rounded corners=2pt,
                minimum size=8mm,fill=green!10,font=\small},
    smR/.style={rectangle,draw,thick,rounded corners=2pt,
                minimum size=8mm,fill=orange!10,font=\small},
    hop/.style={-Latex,thick},
    end/.style={-Latex,thick,dashed},
    label/.style={font=\footnotesize,midway,above}
]
\node[site] (D1) {$D_1$};
\node[site,right=of D1] (D2) {$D_2$};
\node[site,right=of D2] (D3) {$D_3$};
\node[site,right=of D3] (D4) {$D_4$};

\node[below=2mm,font=\footnotesize,gray] at (D1.south) {$q_9\!=\!0$};
\node[below=2mm,font=\footnotesize,gray] at (D2.south) {$q_9\!=\!8$};
\node[below=2mm,font=\footnotesize,gray] at (D3.south) {$q_9\!=\!6$};
\node[below=2mm,font=\footnotesize,gray] at (D4.south) {$q_9\!=\!2$};

\draw[hop] (D1) -- node[label] {$\e^{1/9}$} (D2);
\draw[hop] (D2) -- node[label] {$\e^{2/9}$} (D3);
\draw[hop] (D3) -- node[label] {$\e^{4/9}$} (D4);

\node[smR,left=2.0cm of D1] (dR) {$d_{R,j}$};
\node[smL,right=2.0cm of D4] (qL) {$q_{L,i}$};

\draw[end] (dR) -- node[label] {$\e^{B_j/9}$} (D1);
\draw[end] (D4) -- node[label] {$\e^{A_i/9}$} (qL);

\draw[decorate,decoration={brace,amplitude=5pt,mirror},thick]
  ([yshift=-10mm]D1.south) -- ([yshift=-10mm]D4.south)
  node[midway,below=6pt,font=\footnotesize]
  {$\Sigma = 1+2+4 = 7$ \;(total hop exponent in ninths)};
\end{tikzpicture}
\caption{Schematic of the four-site down-type VLQ chain.
Solid arrows denote nearest-neighbor hops with the indicated
$\e^{h_a/9}$ suppression; dashed arrows show the endpoint
couplings to SM fields.
The $\Z_9$ charges are shown below each site.
The effective Yukawa exponent for entry $(i,j)$ is
$(\Sigma+A_i+B_j)/9$, an exact result of the
chain-inversion theorem (Sec.~\ref{sec:chain-theorem}),
not a perturbative approximation.}
\label{fig:chain}
\end{figure*}

A down-type VLQ chain is introduced with four sites:
\begin{equation}
D_a + \bar{D}_a,\qquad a=1,\dots,4,
\end{equation}
carrying $\Z_9$ charges
\begin{equation}
q_9(D_a) = (0,\,8,\,6,\,2).
\label{eq:D-charges}
\end{equation}
The $Z_2^{\rm (NN)}$ symmetry assigns opposite parities to
adjacent sites, forbidding non-nearest-neighbor mass terms.
The chain structure is shown schematically in
Fig.~\ref{fig:chain}.

The allowed nearest-neighbor hops are
\begin{align}
\bar{D}_1\,D_2\;\Phi &:\quad
\Delta q = 0-8 \equiv 1\!\!\pmod{9},\quad
\text{suppression }\e^{1/9},
\label{eq:hop1}\\
\bar{D}_2\,D_3\;\Phi^2 &:\quad
\Delta q = 8-6 = 2\!\!\pmod{9},\quad
\text{suppression }\e^{2/9},
\label{eq:hop2}\\
\bar{D}_3\,D_4\;\Phi^4 &:\quad
\Delta q = 6-2 = 4\!\!\pmod{9},\quad
\text{suppression }\e^{4/9},
\label{eq:hop3}
\end{align}
giving a discrete hop set $\{1,2,4\}/9$.
Note that the number of flavon insertions at each hop equals
the $\Z_9$ charge difference, ensuring gauge invariance.
The total chain suppression is
$\e^{(1+2+4)/9}=\e^{7/9}$.

\subsection{Endpoint couplings}

The Standard Model (SM) quarks couple to the chain endpoints via
\begin{equation}
\mathcal{L}_{\rm end}
= \lambda_i\,\bar{q}_{L,i}\,\widetilde{H}\,D_4\,
\left(\frac{\Phi}{\Lambda}\right)^{\!A_i}
+ \eta_j\,\bar{D}_1\,d_{R,j}
\left(\frac{\Phi}{\Lambda}\right)^{\!B_j}
+ \text{h.c.},
\label{eq:endpoint}
\end{equation}
where $\widetilde{H}=i\sigma_2 H^*$, and $A_i$, $B_j$ are
integer or rational exponents fixed by the $Z_N$ charges of
the SM fields $q_{L,i}$ and $d_{R,j}$.
The coefficients $\lambda_i$ and $\eta_j$ are $\mathcal{O}(1)$
complex numbers.

\subsection{Up-type chain}
\label{sec:up-chain}

\begin{figure*}[htbp]
\centering
\begin{tikzpicture}[
    node distance=1.6cm,
    dsite/.style={circle,draw,thick,minimum size=9mm,
                  fill=blue!8,font=\small},
    usite/.style={circle,draw,thick,minimum size=9mm,
                  fill=red!8,font=\small},
    smL/.style={rectangle,draw,thick,rounded corners=2pt,
                minimum size=8mm,fill=green!12,font=\small},
    smRd/.style={rectangle,draw,thick,rounded corners=2pt,
                 minimum size=8mm,fill=orange!10,font=\small},
    smRu/.style={rectangle,draw,thick,rounded corners=2pt,
                 minimum size=8mm,fill=purple!10,font=\small},
    hop/.style={-Latex,thick},
    end/.style={-Latex,thick,dashed},
    elabel/.style={font=\scriptsize,midway,above},
    blabel/.style={font=\scriptsize,midway,below}
]
\node[dsite] (D1) {$D_1$};
\node[dsite,right=of D1] (D2) {$D_2$};
\node[dsite,right=of D2] (D3) {$D_3$};
\node[dsite,right=of D3] (D4) {$D_4$};

\draw[hop,blue!70!black] (D1) -- node[elabel] {$\e^{1/9}$} (D2);
\draw[hop,blue!70!black] (D2) -- node[elabel] {$\e^{2/9}$} (D3);
\draw[hop,blue!70!black] (D3) -- node[elabel] {$\e^{4/9}$} (D4);

\node[smRd,left=1.4cm of D1] (dR) {$d_{R,j}$};
\draw[end,blue!70!black] (dR) -- node[elabel] {$\e^{B_j^d/9}$} (D1);

\node[above=2mm,font=\scriptsize,blue!70!black] at (D1.north) {$q_9\!=\!0$};
\node[above=2mm,font=\scriptsize,blue!70!black] at (D2.north) {$q_9\!=\!8$};
\node[above=2mm,font=\scriptsize,blue!70!black] at (D3.north) {$q_9\!=\!6$};
\node[above=2mm,font=\scriptsize,blue!70!black] at (D4.north) {$q_9\!=\!2$};

\node[usite,below=2.5cm of D1] (U1) {$U_1$};
\node[usite,right=of U1] (U2) {$U_2$};
\node[usite,right=of U2] (U3) {$U_3$};
\node[usite,right=of U3] (U4) {$U_4$};

\draw[hop,red!70!black] (U1) -- node[blabel] {$\e^{1/9}$} (U2);
\draw[hop,red!70!black] (U2) -- node[blabel] {$\e^{2/9}$} (U3);
\draw[hop,red!70!black] (U3) -- node[blabel] {$\e^{4/9}$} (U4);

\node[smRu,left=1.4cm of U1] (uR) {$u_{R,j}$};
\draw[end,red!70!black] (uR) -- node[blabel] {$\e^{B_j^u/9}$} (U1);

\node[below=2mm,font=\scriptsize,red!70!black] at (U1.south) {$q_9\!=\!0$};
\node[below=2mm,font=\scriptsize,red!70!black] at (U2.south) {$q_9\!=\!8$};
\node[below=2mm,font=\scriptsize,red!70!black] at (U3.south) {$q_9\!=\!6$};
\node[below=2mm,font=\scriptsize,red!70!black] at (U4.south) {$q_9\!=\!2$};

\node[smL,right=2.0cm of D4,yshift=-1.25cm] (qL) {$q_{L,i}$};

\draw[end,blue!70!black] (D4) -- node[above,sloped,pos=0.4,font=\scriptsize]
  {$\e^{A_i/9}$} (qL);
\draw[end,red!70!black] (U4) -- node[below,sloped,pos=0.4,font=\scriptsize]
  {$\e^{A_i/9}$} (qL);

\node[below=2mm of qL,font=\scriptsize,green!40!black]
  {SU(2)$_L$ doublet};
\node[below=7mm of qL,xshift=2.5cm,font=\small] (ckm)
  {$V_{\rm CKM}=U_{uL}^\dagger U_{dL}$};
\draw[-Latex,thick] (qL) -- (ckm);

\draw[decorate,decoration={brace,amplitude=4pt},thick,blue!70!black]
  ([yshift=7mm]D1.north) -- ([yshift=7mm]D4.north)
  node[midway,above=5pt,font=\scriptsize,blue!70!black]
  {Down chain: $B_j^d=(10,3,0)$};

\draw[decorate,decoration={brace,amplitude=4pt,mirror},thick,red!70!black]
  ([yshift=-7mm]U1.south) -- ([yshift=-7mm]U4.south)
  node[midway,below=5pt,font=\scriptsize,red!70!black]
  {Up chain: $B_j^u=(37,12,0)$};

\end{tikzpicture}
\caption{Combined down-type (blue, upper) and up-type (red, lower) VLQ chains.
Both chains share the same internal hop structure
$\{1,2,4\}/9$ and the same left-handed endpoint dressings
$A_i=(27,18,0)$ [Eq.~\eqref{eq:Ai}], enforced by SU(2)$_L$ gauge invariance
acting on the shared quark doublet $q_{L,i}$.
The right-handed dressings differ [Eq.~\eqref{eq:Bj}]:
$B_j^d=(10,3,0)$ for the down sector and
$B_j^u=(37,12,0)$ for the up sector.
The $\e^{h/9}$ suppressions and the resulting Yukawa
exponents are exact consequences of the upper-triangular
chain mass matrix (Sec.~\ref{sec:convergence}); they do
not rely on $\kappa_a\e^{h_a/9}\ll 1$.
The CKM matrix arises from the mismatch between
the left-handed rotations $U_{dL}$ and $U_{uL}$ that
diagonalize the two Yukawa textures.}
\label{fig:dual-chain}
\end{figure*}
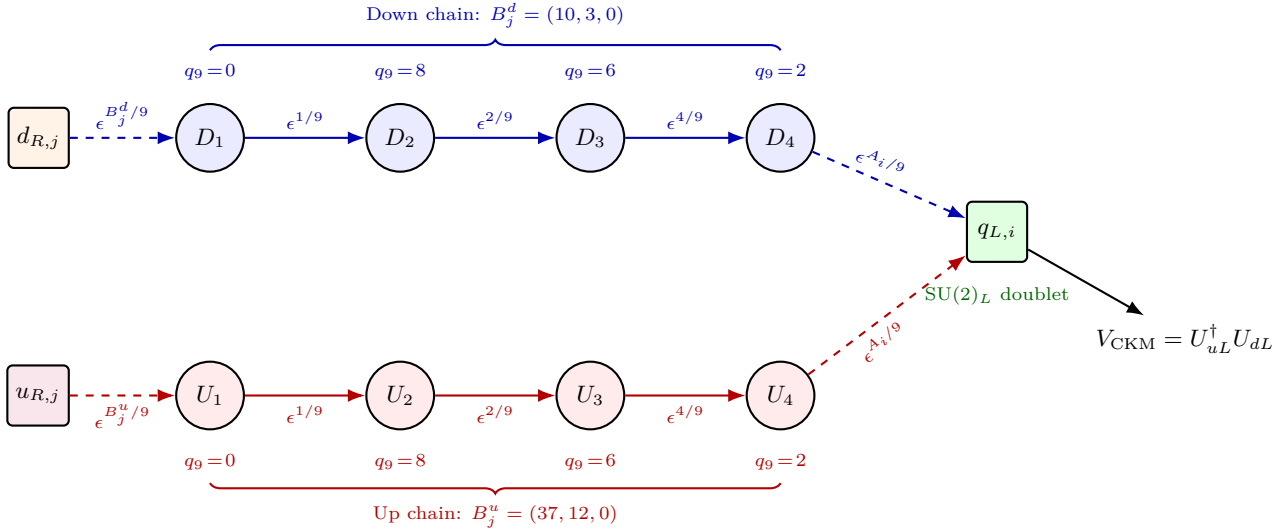

An analogous up-sector chain
$U_a + \bar{U}_a$ ($a=1,\dots,N_u$) generates
\begin{equation}
(Y_u)_{ij} \sim \e^{(\Sigma_u + A_i^u + B_j^u)/9},
\end{equation}
where $\Sigma_u$ is the total hop suppression of the up chain.
The two chains share the same left-handed doublet endpoints,
as shown in Fig.~\ref{fig:dual-chain}.
In a minimal realization, the CKM matrix receives contributions
from both sectors:
\begin{equation}
V_{\rm CKM} = U_{uL}^\dagger\,U_{dL},
\end{equation}
with $U_{uL}$ and $U_{dL}$ the left-handed rotations from
diagonalizing $Y_u Y_u^\dagger$ and $Y_d Y_d^\dagger$
respectively.
In the ``down-dominant'' limit where the up-sector texture is
nearly diagonal, the CKM structure is controlled primarily by
$U_{dL}$, and the analysis of Sec.~\ref{sec:diag} applies
directly.
The full two-chain analysis, which includes CKM interference
between different sectors, is presented in
Appendix~\ref{app:full-diag}.

\subsection{Why four chain sites}

The number of chain sites $N=4$ is not a free choice but is
determined by the requirement that the total chain hop
exponent $\Sigma/9$ be large enough to generate the full
hierarchy between the heaviest and lightest quarks of a
given charge.
For the down sector, the ratio $m_d/m_b\sim\e^{37/9}$
requires a total suppression of
$\e^{37/9}\simeq 10^{-3}$.
The large left-handed charge $Q(Q_1)=3$ and right-handed
charge $Q(d^c_1)=10/9$ together provide the full hierarchy;
the chain hops distribute these charges across nearest-neighbor
links.
A three-site chain with hop set $\{1,2\}/9$ would give
$\Sigma=3$, yielding $\e^{3/9}\simeq 0.57$---insufficient
hierarchy.
A five-site chain would provide more suppression than needed,
at the cost of additional VLQ species.
Four sites is thus the minimal choice consistent with the
observed hierarchy and the discrete hop set $\{1,2,4\}/9$.

\section{General Chain Inversion Theorem}
\label{sec:chain-theorem}

The effective Yukawa coupling between SM quarks $q_{L,i}$
and $d_{R,j}$ is generated by integrating out the heavy
VLQ chain that connects them.
The SM fields do not couple directly to each other;
$q_{L,i}$ couples to the last chain site $D_N$, and
$d_{R,j}$ couples to the first site $D_1$
(Sec.~\ref{sec:field-content}).
The effective coupling is therefore proportional to the
\emph{corner element} $(\mathcal{M}^{-1})_{N1}$ of the
inverse chain mass matrix---the amplitude for a signal
injected at site~$1$ to propagate through the chain and
emerge at site~$N$.

The $Z_2^{(\rm NN)}$ nearest-neighbor symmetry of
Eq.~\eqref{eq:G-symmetry} plays a crucial structural role
here.
By assigning opposite $Z_2$ parities to adjacent chain
sites, it forbids all mass terms between non-adjacent
sites: $\bar{D}_a D_b$ with $|a-b|>1$ is forbidden.
In the $(D_1,\dots,D_N)$ basis, the chain mass matrix
therefore has nonzero entries only on the diagonal
(vectorlike masses $M_a$) and on the first superdiagonal
(nearest-neighbor hops $m_{a,a+1}$), making it
upper-triangular.
This is not a simplifying approximation---it is an exact
consequence of the symmetry.
The payoff is that the inverse of an upper-triangular matrix
is also upper-triangular and can be computed in closed form,
so the corner element and hence the effective Yukawa exponent
are \emph{exact}, with no perturbative expansion required.

\subsection[N-site chain]{$N$-site chain}

Consider an $N$-site chain with upper-triangular mass matrix
in the basis $(D_1,\dots,D_N)$:
\begin{equation}
\mathcal{M} =
\begin{pmatrix}
M_1 & m_{12} & 0 & \cdots & 0 \\
0 & M_2 & m_{23} & \cdots & 0 \\
\vdots & & \ddots & \ddots & \vdots \\
0 & \cdots & 0 & M_{N-1} & m_{N-1,N} \\
0 & \cdots & 0 & 0 & M_N
\end{pmatrix},
\label{eq:M-chain}
\end{equation}
where $M_a$ are the diagonal (vectorlike) masses and
$m_{a,a+1}=\kappa_a M_a\,\e^{h_a/9}$ are the nearest-neighbor
hop masses, with $h_a$ the discrete hop exponents and
$\kappa_a=\mathcal{O}(1)$.

\subsection{Block-inversion formula}

Because $\mathcal{M}$ is upper-triangular, its inverse is
also upper-triangular and can be computed in closed form by
back-substitution~\cite{GolubVanLoan} (see
Appendix~\ref{app:inversion} for the explicit derivation).
The corner element is \emph{exact}:
\begin{equation}
(\mathcal{M}^{-1})_{N1}
= (-1)^{N-1}
\frac{\prod_{k=1}^{N-1} m_{k,k+1}}
     {\prod_{a=1}^{N} M_a},
\label{eq:Minv-general}
\end{equation}
with no series truncation and no requirement that
$m_{a,a+1}\ll M_a$.

For degenerate masses $M_a = M$, this simplifies to
\begin{equation}
\boxed{
(\mathcal{M}^{-1})_{N1}
= \frac{(-1)^{N-1}}{M}\prod_{k=1}^{N-1}\frac{m_{k,k+1}}{M}
= \frac{(-1)^{N-1}\prod_k\kappa_k}{M}\,
\e^{\,\Sigma/9}
}
\label{eq:Minv-degen}
\end{equation}
where
$\Sigma\equiv\sum_{k=1}^{N-1}h_k$
is the total hop exponent and the overall sign
is absorbed into the Yukawa coupling.
Crucially, this formula holds for \emph{any} values
of $\kappa_a$ and $\e^{h_a/9}$---it is an algebraic
identity, not the leading term of a perturbative
expansion.

\subsection[Explicit 4-site example]{Explicit $4$-site example}

For the down chain of Sec.~\ref{sec:field-content},
the mass matrix is
\begin{equation}
\mathcal{M}_D =
\begin{pmatrix}
M & \kappa_1 M\e^{1/9} & 0 & 0 \\
0 & M & \kappa_2 M\e^{2/9} & 0 \\
0 & 0 & M & \kappa_3 M\e^{4/9} \\
0 & 0 & 0 & M
\end{pmatrix}.
\label{eq:MD-explicit}
\end{equation}
The block-inversion formula~\eqref{eq:Minv-degen} gives
\begin{equation}
(\mathcal{M}_D^{-1})_{41}
= -\frac{\kappa_1\kappa_2\kappa_3}{M}\,
\e^{(1+2+4)/9}
= -\frac{\kappa_1\kappa_2\kappa_3}{M}\,\e^{7/9},
\label{eq:MD-corner}
\end{equation}
where the sign is absorbed into the Yukawa coupling.
This result is \emph{exact}: numerical evaluation for
$\kappa_a=1$ gives
$|(\mathcal{M}_D^{-1})_{41}|=\e^{7/9}/M$ to machine
precision, and the agreement persists even for
$\kappa_a\gg 1$.

\subsection{Path-Sum Theorem}

Combining the chain corner element~\eqref{eq:MD-corner}
with the endpoint couplings~\eqref{eq:endpoint}, the
effective Yukawa matrix is
\begin{equation}
(Y_d)_{ij}
= \lambda_i\eta_j\,
\frac{\kappa_1\kappa_2\kappa_3}{M}\,
\e^{(\Sigma + A_i + B_j)/9},
\label{eq:Yd-pathsum}
\end{equation}
with $\Sigma=7$.

\begin{center}
\fbox{
\parbox{0.92\linewidth}{
\textbf{Path-Sum Theorem.}
In a nearest-neighbor lattice chain, the effective Yukawa
exponent for each entry $(i,j)$ equals the sum of the
discrete hop exponents $\Sigma=\sum h_k$ plus the endpoint
dressings $A_i$ and $B_j$:
\begin{equation}
p_{ij} = \frac{\Sigma + A_i + B_j}{9}.
\end{equation}
This realizes the $\Bpar$-lattice exponent algebra
dynamically.
}}
\end{center}

\subsection{Convergence of the chain expansion}
\label{sec:convergence}

A potential concern is that the individual hop
parameters $\kappa_a\,\e^{h_a/9}$ are not uniformly
small---the smallest hop has $h_1=1$, giving
$\e^{1/9}\simeq 0.83$, so that for $\kappa_1\sim 1$
the ratio $m_{12}/M\sim 0.83$ is not a conventional
small parameter.
We now show that this does \emph{not} compromise
the framework, because the quantity
$\kappa_a\e^{h_a/9}$ is never used as a perturbative
expansion parameter.

\paragraph{The chain inversion is exact.}
The corner element
formula~\eqref{eq:Minv-general}--\eqref{eq:Minv-degen}
is an algebraic identity for upper-triangular matrices,
derived by back-substitution with no series truncation
(see Appendix~\ref{app:inversion}).
As verified numerically,
the formula reproduces the
exact matrix inverse to machine precision for all
values of $\kappa_a$, including
$\kappa_a\,\e^{h_a/9}\gg 1$.
The $Z_2^{(\rm NN)}$ locality symmetry enforces the
upper-triangular structure, and the Path-Sum Theorem
therefore holds without any small-parameter assumption.

\paragraph{The Yukawa diagonalization uses different
expansion parameters.}
The perturbative diagonalization of the effective
$3\times 3$ Yukawa matrix
(Sec.~\ref{sec:diag} and Appendix~\ref{app:full-diag})
expands in the left-handed mixing angles
$\theta_{ij}^{dL}\sim\e^{|Q(Q_i)-Q(Q_j)|}$,
whose magnitudes are
\begin{equation}
\theta_{12}^{dL}\sim\e\simeq 0.19,\quad
\theta_{23}^{dL}\sim\e^2\simeq 0.035,\quad
\theta_{13}^{dL}\sim\e^3\simeq 0.007.
\label{eq:diag-params}
\end{equation}
These are the physical expansion parameters of the
diagonalization, and all are comfortably small.
The quantity $\e^{1/9}\simeq 0.83$ does not appear as
an expansion parameter at this stage.

\paragraph[What kappa*epsilon O(1) does affect.]{What $\kappa_a\e^{h_a/9}\sim\mathcal{O}(1)$
does affect.}
The chain mass eigenstates receive
$\mathcal{O}(\kappa^2\e^{2h_a/9})$ shifts relative to
the site-basis states (see Sec.~\ref{sec:collider}),
and these splittings can be large:
$\sim 69\%$ for the first hop ($h_1=1$) and
$\sim 47\%$ for the second ($h_2=2$).
These corrections redistribute mass among the VLQ
eigenstates but do not alter the exponent structure
of the effective Yukawa.
The VLQ spectrum and SM--VLQ mixing angles
(Sec.~\ref{sec:flavor-constraints}) should therefore
be understood as order-of-magnitude estimates,
with $\mathcal{O}(1)$ corrections from the non-perturbative
regime of the chain splittings.

\paragraph[Robustness under O(1) variation.]{Robustness under $\mathcal{O}(1)$ variation.}
Exact numerical diagonalization of the $3\times 3$ down-type Yukawa
for $10^3$ random $\mathcal{O}(1)$ complex coefficients
$c^d_{ij}$ with $|c^d_{ij}|\in[0.5,\,2.0]$ confirms that the
eigenvalue ratios
$y_d/\e^{37/9}$, $y_s/\e^{7/3}$, and $y_b/1$ remain
$\mathcal{O}(1)$ in every case, with means of
$2.2$, $1.6$, and $1.5$ respectively.
The $\e$-scaling is therefore robust: higher-order corrections
modify only the $\mathcal{O}(1)$ prefactors.

\section[Connection to the B-Lattice Quark Textures]{Connection to the $\Bpar$-Lattice Quark Textures}
\label{sec:Blattice-connection}

The $\Bpar$-lattice quark
program~\cite{Barger2025bfn,Barger2025bfnb,LatticeFlavonQuarkMixing}
established the Yukawa exponent matrices
\begin{equation}
\begin{aligned}
p^{u}_{ij} &=
{\setlength{\arraycolsep}{6pt}\renewcommand{\arraystretch}{1.15}
\begin{pmatrix}
\tfrac{64}{9} & \tfrac{13}{3} & 3\\[2pt]
\tfrac{55}{9} & \tfrac{10}{3} & 2\\[2pt]
\tfrac{37}{9} & \tfrac{4}{3}  & 0
\end{pmatrix}},\\[8pt]
p^{d}_{ij} &=
{\setlength{\arraycolsep}{6pt}\renewcommand{\arraystretch}{1.15}
\begin{pmatrix}
\tfrac{37}{9} & \tfrac{10}{3} & 3\\[2pt]
\tfrac{28}{9} & \tfrac{7}{3}  & 2\\[2pt]
\tfrac{10}{9} & \tfrac{1}{3}  & 0
\end{pmatrix}}.
\end{aligned}
\label{eq:pud-Blattice}
\end{equation}
These factorize as $p^f_{ij}=Q(Q_i)+Q(f^c_j)$ with shared
left-handed doublet charges $Q(Q_i)=(3,2,0)$ and
right-handed charges $Q(u^c_j)=(\tfrac{37}{9},\tfrac{4}{3},0)$,
$Q(d^c_j)=(\tfrac{10}{9},\tfrac{1}{3},0)$.

These charges are \emph{not} free parameters chosen to fit
the data; they are determined bottom-up from the observed
mass spectrum in a three-step sequence that leaves no
free choices.

First, the value $\Bpar=75/14$ is itself fixed
by the requirement that the six quark mass ratios
$m_f/m_{f_3}=\e^{p_f}$ simultaneously yield
$\mathcal{O}(1)$ coefficients
$c_f\simeq 1.00$~\cite{Barger2025bfn}.
No other value of $\Bpar$ reproduces all six
ratios to this precision
(Sec.~\ref{sec:numerical-benchmark}).

Second, once the diagonal exponents $n_f$ are read off
from the measured quark masses at $M_Z$, the
factorization $p^f_{ij}=Q(Q_i)+Q(f^c_j)$ is an
\emph{overdetermined} system: $9$ entries per sector
are constrained by $3+3$ charges, so $12$ independent
exponents must accommodate $18$ texture entries across
both sectors.
With the convention $Q(Q_3)=Q(f^c_3)=0$ (third
generation carries zero charge), the decomposition is
\emph{unique}: $Q(Q_i)=p^d_{i3}$ and
$Q(f^c_j)=p^f_{3j}$, read directly from the third row
and column of the exponent matrix.
That the resulting charges are rational
multiples of $1/9$ is a consequence of the observed mass
ratios, not a design choice.

Third, the factorization succeeds---with all
$\mathcal{O}(1)$ coefficients essentially unity
(Sec.~\ref{sec:numerical-benchmark})---which is a
nontrivial check, not a parametric fit.
The subsequent verification that these
phenomenologically determined charges satisfy the discrete
anomaly conditions (Sec.~\ref{sec:anomaly}) is therefore
a prediction of the framework, not an input.

The Yukawa matrices are
\begin{equation}
(Y_f)_{ij} = c^f_{ij}\,\e^{\,p^f_{ij}},
\qquad f = u,\,d,
\label{eq:Yf-def}
\end{equation}
where the $c^f_{ij}$ are $\mathcal{O}(1)$ complex coefficients
encoding the product of vertex couplings along the chain.

In the multi-messenger realization of
Ref.~\cite{Barger2025bfnb}, each Yukawa entry receives
contributions from several chain configurations (messenger
species), so that the effective coefficient takes the form
\begin{equation}
C^{f}_{ij}
= 1 + e^{i\phi_f}\,\e^{\Delta^f_{ij}}
    + e^{i\psi_f}\,\e^{\Delta'^f_{ij}},
\label{eq:Ceff}
\end{equation}
where $\Delta^f_{ij},\Delta'^f_{ij}\geq 0$ are additional
suppression exponents from the subleading chains, and
$\phi_f$, $\psi_f$ are relative phases.
The terminology ``multi-messenger'' is borrowed, with intent,
from astrophysics: just as multi-messenger astronomy
combines independent observation channels---gravitational
waves, neutrinos, photons---to reconstruct a single
astrophysical event, multi-messenger flavor physics combines
independent heavy-fermion chains to construct each Yukawa
coupling, with the interference pattern encoding the
CP-violating phase.
The leading term is unity, and the corrections are
suppressed by positive powers of $\e$, so
$|C^f_{ij}|$ is generically $\mathcal{O}(1)$ without
fine tuning.
The phases $\phi_f$ and $\psi_f$ are the physical origin
of CP violation: they generate the CKM phase through the
adjacent-exponent interference mechanism discussed in
Sec.~\ref{sec:CP-phase}.
The $\mathcal{O}(1)$ coefficients $c^f_{ij}$ in
Eq.~\eqref{eq:Yf-def} are identified with these effective
coefficients, $c^f_{ij}=C^f_{ij}\times(\text{overall
normalization})$.

For $\e\ll 1$, the hierarchy is controlled by the exponent
matrix $p^f$; the coefficients provide $\mathcal{O}(1)$
modulation but do not alter the power-counting.

Each Yukawa matrix is diagonalized by a bidiagonal
(singular-value) decomposition:
\begin{equation}
U_{fL}^\dagger\, Y_f\, U_{fR}
= Y_f^{\rm diag}
\equiv \diag(y_{f_1},\,y_{f_2},\,y_{f_3}),
\label{eq:Y-diag-def}
\end{equation}
where $U_{fL}$ and $U_{fR}$ are unitary matrices acting on
the left-handed and right-handed fields respectively,
and the $y_{f_i}$ are the real, non-negative Yukawa eigenvalues
related to the quark masses by
$m_{f_i} = y_{f_i}\,v/\sqrt{2}$.
The CKM matrix is then
\begin{equation}
V_{\rm CKM} = U_{uL}^\dagger\,U_{dL}.
\label{eq:VCKM-def}
\end{equation}

The quark mass hierarchies are
\begin{equation}
m_u : m_c : m_t \sim \e^{64/9} : \e^{10/3} : 1,
\end{equation}
\begin{equation}
m_d : m_s : m_b \sim \e^{37/9} : \e^{7/3} : 1,
\end{equation}
and the CKM mixing angles scale as
\begin{align}
|V_{us}| &\sim \e^{8/9} \simeq 0.22, \label{eq:Vus}\\
|V_{cb}| &\sim \e^{17/9} \simeq 0.042, \label{eq:Vcb}\\
|V_{ub}| &\sim \e^{10/3} \simeq 0.0035, \label{eq:Vub}
\end{align}
in excellent agreement with the Particle Data Group (PDG)
values~\cite{PDG2024}.

In the chain framework, these exponents are reproduced by
the path-sum formula
\begin{equation}
p^f_{ij} = \frac{A_i + B_j^f}{9},
\end{equation}
where $A_i$ are the shared left-handed endpoint dressings
(in ninths) and $B_j^f$ are the right-handed dressings.
SU(2)$_L$ gauge invariance requires the same $A_i$ for both
sectors:
\begin{equation}
A_i = (27,\,18,\,0),
\label{eq:Ai}
\end{equation}
with
\begin{equation}
B_j^d = (10,\,3,\,0),\qquad
B_j^u = (37,\,12,\,0).
\label{eq:Bj}
\end{equation}
The chain hop contributions are absorbed into the endpoint
dressings, so the factorized charge structure
$p^f_{ij}=Q(Q_i)+Q(f^c_j)$ is manifest.
This demonstrates that the phenomenologically successful
$\Bpar$-lattice textures emerge naturally from the
discrete charge assignments of
Sec.~\ref{sec:field-content}.

\subsection{Numerical benchmark}
\label{sec:numerical-benchmark}

With $\e=14/75\simeq 0.1867$ and $\overline{\rm MS}$ quark
masses at $M_Z$~\cite{Barger2025bfn,PDG2024}
($m_u=1.11$~MeV, $m_d=2.82$~MeV, $m_s=55.7$~MeV,
$m_c=0.629$~GeV, $m_b=2.794$~GeV, $m_t=169$~GeV),
the $\Bpar$-lattice gives:
\begin{equation}
\frac{m_d}{m_b}\simeq \e^{37/9}\cdot c_d
= 1.01\times 10^{-3}\cdot c_d,
\end{equation}
\begin{equation}
\frac{m_s}{m_b}\simeq \e^{7/3}\cdot c_s
= 1.99\times 10^{-2}\cdot c_s,
\end{equation}
\begin{equation}
\frac{m_u}{m_t}\simeq \e^{64/9}\cdot c_u
= 6.6\times 10^{-6}\cdot c_u,
\end{equation}
\begin{equation}
\frac{m_c}{m_t}\simeq \e^{10/3}\cdot c_c
= 3.72\times 10^{-3}\cdot c_c,
\end{equation}
with $c_d\simeq c_s\simeq c_u\simeq c_c\simeq 1.00$.
All four $\mathcal{O}(1)$ coefficients are essentially
unity---the $\Bpar$-lattice exponents of
Eq.~\eqref{eq:pud-Blattice} reproduce the physical quark
mass ratios $m_d/m_b\simeq 1.01\times 10^{-3}$,
$m_s/m_b\simeq 1.99\times 10^{-2}$,
$m_u/m_t\simeq 6.6\times 10^{-6}$, and
$m_c/m_t\simeq 3.7\times 10^{-3}$
with no fine tuning.

\section{Perturbative Diagonalization of the Down-Type Texture}
\label{sec:diag}

We illustrate the sequential diagonalization procedure
using the physical down-type texture of
Eq.~\eqref{eq:pud-Blattice}.
Setting all $\mathcal{O}(1)$ coefficients to unity for
transparency, the Yukawa matrix is
\begin{equation}
Y_d \sim
\begin{pmatrix}
\e^{37/9} & \e^{10/3} & \e^{3} \\[3pt]
\e^{28/9} & \e^{7/3}  & \e^{2} \\[3pt]
\e^{10/9} & \e^{1/3}  & 1
\end{pmatrix}.
\label{eq:Yd-physical}
\end{equation}
Because $p^d_{ij}=Q(Q_i)+Q(d^c_j)$ factorizes, the matrix
is \emph{not} symmetric: the left-handed charges
$Q(Q_i)=(3,2,0)$ differ from the right-handed charges
$Q(d^c_j)=(\tfrac{10}{9},\tfrac{1}{3},0)$.
The diagonalization therefore requires separate left- and
right-handed rotations,
$U_{dL}^\dagger\,Y_d\,U_{dR}=Y_d^{\rm diag}$.

\subsection{Third generation}

The $(3,3)$ element dominates the matrix
($p^d_{33}=0$, i.e.\ unsuppressed):
\begin{equation}
y_b \simeq 1.
\end{equation}
The physical $b$-quark mass is set by the overall Yukawa
normalization $m_b = y_b\,v/\sqrt{2}$.
In the companion analysis of Ref.~\cite{Barger2025bfn},
this is anchored to the charged-lepton sector through
$m_b^{\rm ref}=\varphi\,m_\tau^{\overline{\rm MS}}(M_Z)$,
where $\varphi=(1+\sqrt{5})/2$ is the golden ratio---a
Georgi--Jarlskog-style relation that fixes the reference
scale from which all other quark masses follow via
$m_f=m_b^{\rm ref}\cdot\Bpar^{\,n_f}$.

\subsection[2--3 rotation]{$2$--$3$ rotation}

The $(2,3)$ element is removed by a left-handed rotation
with angle
\begin{equation}
\theta_{23}^{dL}
\approx \frac{|(Y_d)_{23}|}{|(Y_d)_{33}|}
= \frac{\e^{2}}{1}
= \e^{2}\simeq 0.035.
\label{eq:t23-phys}
\end{equation}
This is parametrically $\e^{Q(Q_2)-Q(Q_3)}=\e^{2}$.
The right-handed rotation has a different angle,
$\theta_{23}^{dR}\approx\e^{1/3}\simeq 0.57$,
reflecting the asymmetry between left and right charges.

\subsection{Second generation}

After decoupling the third generation, the effective
$(2,2)$ element is
\begin{equation}
y_s \simeq \e^{7/3}
- \frac{\e^{2}\cdot\e^{1/3}}{1}
= \e^{7/3}(1 - \e^0) + \cdots,
\end{equation}
which at leading order gives
\begin{equation}
y_s \sim \e^{7/3}\simeq 0.020,
\end{equation}
since the seesaw correction from the $2$--$3$ block
involves $\e^{2+1/3}=\e^{7/3}$, the same power as the
diagonal element itself.
The physical coefficient thus depends on the
$\mathcal{O}(1)$ factors; generically
$y_s = |c_{22}^d - c_{23}^d c_{32}^d/c_{33}^d|\,\e^{7/3}$.

\subsection[1--2 rotation]{$1$--$2$ rotation}

The residual $(1,2)$ element after the $2$--$3$ rotation
scales as $\e^{10/3}$.
The $1$--$2$ left-handed mixing angle is therefore
\begin{equation}
\theta_{12}^{dL}
\approx \frac{\e^{10/3}}{\e^{7/3}}
= \e^{10/3-7/3} = \e\simeq 0.187.
\label{eq:t12-phys}
\end{equation}
This is $\e^{Q(Q_1)-Q(Q_2)}=\e^{3-2}=\e$.

\subsection[First generation and 1--3 mixing]{First generation and $1$--$3$ mixing}

The lightest eigenvalue emerges as
\begin{equation}
y_d \sim \e^{37/9}\simeq 1.0\times 10^{-3}.
\end{equation}
The direct $1$--$3$ mixing angle is
\begin{equation}
\theta_{13}^{dL}
\approx \frac{\e^{3}}{1}
= \e^{3}\simeq 0.007.
\label{eq:t13-phys}
\end{equation}
This is $\e^{Q(Q_1)-Q(Q_3)}=\e^{3}$.

\subsection{Summary of eigenvalue and mixing hierarchy}

Collecting the results for the down sector:
\begin{align}
y_b &\sim 1, &
y_s &\sim \e^{7/3}, &
y_d &\sim \e^{37/9},
\label{eq:masses-phys}\\[4pt]
\theta_{23}^{dL} &\sim \e^{2}, &
\theta_{12}^{dL} &\sim \e, &
\theta_{13}^{dL} &\sim \e^{3}.
\label{eq:angles-phys}
\end{align}
Each left-handed mixing angle is controlled by the
doublet charge difference
$\theta_{ij}^{dL}\sim\e^{|Q(Q_i)-Q(Q_j)|}$.

The right-handed rotation matrix $U_{dR}$ is determined
by the right-handed charge differences
$Q(d^c_i)-Q(d^c_j)$:
\begin{align}
\theta_{23}^{dR} &\sim \e^{1/3}\simeq 0.57, &
\theta_{12}^{dR} &\sim \e^{7/9}\simeq 0.27, \nonumber\\
\theta_{13}^{dR} &\sim \e^{10/9}\simeq 0.15.
\label{eq:angles-dR}
\end{align}
Since the down-type right-handed charges
$Q(d^c_j)=(\tfrac{10}{9},\tfrac{1}{3},0)$ are more
closely spaced than the left-handed charges
$Q(Q_i)=(3,2,0)$, the right-handed rotations are
generically larger than the left-handed ones.
The right-handed rotations do not enter the CKM matrix
but are relevant for right-handed currents in
extensions beyond the SM.

For the up sector, the same logic gives
$\theta_{ij}^{uL}\sim\theta_{ij}^{dL}$ (shared doublet
charges), while the right-handed angles are
\begin{align}
\theta_{23}^{uR} &\sim \e^{4/3}\simeq 0.11, &
\theta_{12}^{uR} &\sim \e^{25/9}\simeq 0.009, \nonumber\\
\theta_{13}^{uR} &\sim \e^{37/9}\simeq 0.001.
\label{eq:angles-uR}
\end{align}
The up-type right-handed charges
$Q(u^c_j)=(\tfrac{37}{9},\tfrac{4}{3},0)$ are more
widely spaced, yielding smaller (more hierarchical)
right-handed rotations than in the down sector.
Because the up sector shares the same doublet charges
$Q(Q_i)=(3,2,0)$, the up-type left-handed angles
have \emph{identical} $\e$-scaling:
$\theta_{ij}^{uL}\sim\theta_{ij}^{dL}$.
The CKM elements therefore arise from the interference
$V_{ij}\sim\theta_{ij}^{dL}-\theta_{ij}^{uL}\,e^{i\phi}$,
as derived in Appendix~\ref{app:full-diag}.

A key feature of the non-symmetric texture is that
$\theta_{12}^{dL}\sim\e\neq\theta_{23}^{dL}\sim\e^2$,
so the physical hierarchy
$|V_{us}|\gg|V_{cb}|$
is built into the charge assignments.
This resolves the degeneracy
$\theta_{12}\sim\theta_{23}$ that would appear in a
simplified symmetric texture with a common charge
difference.

\subsection{Jarlskog invariant scaling}

The Jarlskog invariant~\cite{Jarlskog} scales as
\begin{equation}
J \sim \theta_{12}\,\theta_{23}\,\theta_{13}\,\sin\delta.
\end{equation}
Using the full $\Bpar$-lattice
scalings~\eqref{eq:Vus}--\eqref{eq:Vub}:
\begin{equation}
J \sim \e^{8/9}\cdot\e^{17/9}\cdot\e^{10/3}\cdot\sin\delta
\sim \e^{55/9}\,\sin\delta
\simeq 3.5\times 10^{-5}\sin\delta,
\label{eq:J-scaling}
\end{equation}
consistent with the PDG value
$J=(3.00\pm0.13)\times 10^{-5}$~\cite{PDG2024}
for $\sin\delta\sim\mathcal{O}(1)$.

\section{CKM Phase from Adjacent Exponent Interference}
\label{sec:CP-phase}

Physical CP violation arises from interference between
Yukawa entries whose exponents differ by $\mathcal{O}(1/9)$.
To see this, introduce a single complex endpoint coefficient
in the $2$--$3$ entry:
\begin{equation}
Y_{23} = c_1\,\e^{4/9} + c_2\,\e^{5/9}\,e^{i\alpha},
\end{equation}
where $c_1,c_2=\mathcal{O}(1)$ are real and $\alpha$ is a
physical phase originating from the complex flavon VEV.
The CKM phase is then
\begin{equation}
\delta_{\rm CKM}
\simeq
\arg\!\left(1 + r\,\e^{1/9}\,e^{i\alpha}\right),
\label{eq:delta-CKM}
\end{equation}
with $r=c_2/c_1$.
Since $\e^{1/9}\simeq 0.83$, the interference is
$\mathcal{O}(1)$ and generates a large CP phase without
fine tuning.
(Note that $\e^{1/9}$ being $\mathcal{O}(1)$ is a
\emph{virtue} for CP violation; as shown in
Sec.~\ref{sec:convergence}, it does not compromise
the chain inversion or the Yukawa diagonalization.)
This is the same mechanism identified in
Paper~III~\cite{LatticeFlavonQuarkMixing} and is a
direct consequence of the densely spaced ninths lattice.

\subsection{Structural down-dominance}
\label{sec:down-dominance}

A natural question is whether the
CP-violating phase requires a
specific relationship between the up- and down-sector
multi-messenger phases $\phi_f$, $\psi_f$.
We now show that the CKM phase is
\emph{structurally controlled by the down sector},
so that the up-sector phases are largely irrelevant.

The Fritzsch--Xing decomposition
(Appendix~\ref{app:full-diag}) expresses the Cabibbo
angle as
\begin{equation}
|V_{us}| \simeq |s_d - s_u\,e^{i\phi_{\rm FX}}|,
\end{equation}
with $s_d\sim\e^{8/9}\simeq 0.22$ and
$s_u\sim\e^{17/9}\simeq 0.042$.
The ratio
\begin{equation}
\frac{s_u}{s_d} = \e^{17/9-8/9} = \e\simeq 0.19
\label{eq:down-dominance-ratio}
\end{equation}
quantifies the structural suppression:
the up-sector contribution enters $|V_{us}|$ at the
$\sim\!19\%$ level, regardless of the FX phase
$\phi_{\rm FX}$.
Varying $\phi_{\rm FX}$ over its full range changes
$|V_{us}|$ by only $\pm s_u\simeq 0.04$ around the
down-sector baseline $s_d\simeq 0.22$.
For $|V_{cb}|$ the suppression is even stronger:
the up-sector correction enters at order
$\e^{17/18}\simeq 0.20$ relative to the down-sector
contribution.

The physical origin of this hierarchy is the
asymmetry in right-handed charge spacings:
\begin{equation}
Q(d^c_j) = \bigl(\tfrac{10}{9},\;\tfrac{1}{3},\;0\bigr)
\quad\text{vs.}\quad
Q(u^c_j) = \bigl(\tfrac{37}{9},\;\tfrac{4}{3},\;0\bigr).
\end{equation}
The wider spacing of the up-sector charges produces
a more hierarchical up-type Yukawa, so that the
up-sector eigenvalue ratios $m_u/m_c$ and $m_c/m_t$
are much smaller than their down-sector counterparts
$m_d/m_s$ and $m_s/m_b$.
The CKM mixing angles---being differences of left-handed
rotations from the two sectors---are therefore dominated
by whichever sector has the larger eigenvalue ratios,
which is structurally the down sector.

The benchmark phases of Appendix~\ref{app:chain-examples}
(vanishing $\phi_u$, $\psi_u$; near-$\pi$ values
of $\phi_d$, $\psi_d$) are therefore an illustrative
simplification, not a fine-tuned choice.
The vanishing up-sector phases can be replaced by
arbitrary $\mathcal{O}(1)$ values with only a
$\sim\!19\%$ change in $|V_{us}|$ and a correspondingly
small shift in the Jarlskog invariant.

\subsection[Genericity of O(1) CP violation]{Genericity of $\mathcal{O}(1)$ CP violation}
\label{sec:CP-genericity}

To quantify the extent to which large CP violation
is a generic prediction of the lattice structure
(rather than a consequence of specific phase choices),
we perform a numerical scan over random
$\mathcal{O}(1)$ complex coefficients
$c^f_{ij}$ with $|c^f_{ij}|\in[0.5,\,2.0]$ and
uniformly distributed phases, independently for both
sectors.
For each of $5\times 10^3$ trials, the full Yukawa
matrices~\eqref{eq:Yf-def} are constructed and the CKM
matrix is extracted by exact singular-value decomposition.

The results (Table~\ref{tab:CP-scan}) demonstrate that:
\begin{enumerate}
\item The Jarlskog invariant falls within a factor of~3
of the PDG value
$J=(3.00\pm 0.13)\times 10^{-5}$~\cite{PDG2024}
in $55\%$ of all trials, and within the range
$[10^{-6},\,10^{-3}]$ in $98\%$ of trials.
\item The effective $|\!\sin\delta|$ exceeds $0.3$ in
$81\%$ of trials---large CP violation is the generic
outcome.
\item The CKM mixing magnitudes
$|V_{us}|$, $|V_{cb}|$, and $|V_{ub}|$ cluster around
their PDG values with $\mathcal{O}(1)$ scatter.
\end{enumerate}

\begin{table*}[htbp]
\centering
\caption{CKM observables from $5\times 10^3$ random
$\mathcal{O}(1)$ complex coefficient trials.
``Median'' and ``IQR'' denote the median and
interquartile range.}
\label{tab:CP-scan}
\setlength{\tabcolsep}{3pt}
\renewcommand{\arraystretch}{1.10}
\begin{tabular}{lccc}
\toprule
Observable & Median & IQR & PDG \\
\midrule
$J/10^{-5}$ & 4.1 & $[1.2,\;10]$ & $3.0\pm0.1$ \\
$|V_{us}|$  & 0.26  & $[0.16,\;0.39]$  & 0.225 \\
$|V_{cb}|$  & 0.042 & $[0.025,\;0.060]$ & 0.042 \\
$|V_{ub}|$  & 0.010 & $[0.004,\;0.017]$ & 0.004 \\
\bottomrule
\end{tabular}
\end{table*}

\noindent
\emph{Phase counting.}
The standard phase-counting theorem for three generations
gives $(n-1)(n-2)/2=1$ physical CP-violating
phase~\cite{Jarlskog}.
The multi-messenger phases $\phi_f$, $\psi_f$ are not
additional parameters: they are specific combinations
of the UV coupling phases that survive after all field
rephasing.
The chain framework therefore has the same number of
irreducible phase inputs as the SM---one.
Its advantage is that the lattice exponent algebra
determines the \emph{scaling} $J\sim\e^{55/9}\sin\delta$
structurally, while the dense ninths spacing ensures
that the remaining free parameter $\sin\delta$ is
generically $\mathcal{O}(1)$.

\section[Flavor Constraints: Delta F=2 and FCNCs]{Flavor Constraints: $\Delta F=2$ and FCNCs}
\label{sec:flavor-constraints}

\subsection{SM--VLQ mixing}

Integrating out the heavy chain states induces mixing between
SM quarks and VLQs.
The left-handed mixing angles scale as
\begin{equation}
\theta_L^i \sim \lambda_i\,\frac{v}{M}\,\e^{n_i/9},
\label{eq:thetaL}
\end{equation}
where $v\simeq 174$~GeV is the electroweak VEV, $M$ is the
common VLQ mass, $n_i$ encodes the total suppression for
generation $i$, and $\lambda_i$ is the $\mathcal{O}(1)$
endpoint coupling from Eq.~\eqref{eq:endpoint}.

For a benchmark mass $M=2$~TeV, $\lambda_i\sim 1$, and
the $\Bpar$-lattice charges, the angles are
\begin{align}
\theta_L^b &\sim \lambda_3\,\frac{v}{M}\,\e^{0}
\sim 10^{-1}\,\lambda_3,
\label{eq:tL-b}\\
\theta_L^s &\sim \lambda_2\,\frac{v}{M}\,\e^{7/3}
\sim 2\times 10^{-3}\,\lambda_2,
\label{eq:tL-s}\\
\theta_L^d &\sim \lambda_1\,\frac{v}{M}\,\e^{37/9}
\sim 10^{-4}\,\lambda_1.
\label{eq:tL-d}
\end{align}
The precise value of $\lambda_3$ is constrained by
the $Zb\bar{b}$ vertex measurement (see
Sec.~\ref{sec:Z-FCNC}); all other angles carry
additional $\e$-suppression from the lattice hierarchy.

\subsection[Delta F=2 operators]{$\Delta F=2$ operators}

The dominant new-physics contribution to meson mixing arises
from tree-level VLQ exchange, generating four-fermion operators
suppressed by
\begin{equation}
C_{\Delta F=2}
\sim \frac{(\theta_L^i)^2(\theta_L^j)^2}{M^2}.
\label{eq:DF2}
\end{equation}

For $K$--$\bar{K}$ mixing ($i=d$, $j=s$):
\begin{equation}
C_{K\bar K}
\sim \frac{(10^{-4})^2(2\times 10^{-3})^2}{(2~\text{TeV})^2}
\sim 10^{-18}~\text{GeV}^{-2},
\end{equation}
well below the experimental bound
$C_K^{\rm exp}\lesssim 10^{-12}~\text{GeV}^{-2}$~\cite{UTfit}.

For $B_s$--$\bar{B}_s$ mixing ($i=s$, $j=b$):
\begin{equation}
C_{B_s}
\sim \frac{(2\times 10^{-3})^2(10^{-1})^2}{(2~\text{TeV})^2}
\sim 10^{-14}~\text{GeV}^{-2},
\end{equation}
also consistent with current constraints.

The key point is that the $\Bpar$-lattice hierarchy in the
mixing angles~\eqref{eq:tL-b}--\eqref{eq:tL-d} provides a
\emph{built-in Glashow--Iliopoulos--Maiani (GIM)-like
suppression}~\cite{GIM} of FCNCs---the same
lattice that generates the mass hierarchy simultaneously
protects against flavor violation.

\subsection{One-loop dipole operators}
\label{sec:dipole}

The electromagnetic dipole Wilson coefficient $C_7$,
which governs $b\to s\gamma$, receives three types of
VLQ contributions:

\emph{(i)~Modified $Wtb$ vertex.}
The SM--VLQ mixing shifts the $Wtb$ coupling by
$\delta g_{Wtb}/g_{Wtb}\sim -(\theta_L^b)^2/2$,
modifying the dominant SM one-loop diagram.
The fractional shift in $C_7$ is
\begin{equation}
\frac{\delta C_7}{C_7^{\rm SM}}
\sim -(\theta_L^b)^2\sim -3\times 10^{-3},
\end{equation}
two orders of magnitude below the experimental
sensitivity $|\delta C_7/C_7^{\rm SM}|\lesssim 0.10$
from the measured ${\rm BR}(B\to X_s\gamma)$~\cite{PDG2024}.

\emph{(ii)~$W$--VLQ loop.}
A new penguin diagram with a VLQ running in the loop
alongside the $W$ boson contributes
\begin{equation}
C_7^{\rm VLQ}
\sim \frac{\alpha_W}{4\pi}\,
\frac{\theta_L^b\,\theta_L^s}{V_{tb}\,V_{ts}}\,
f_7\!\left(\frac{m_t^2}{M^2}\right),
\end{equation}
where $f_7(x)\to 7x/36$ for $x\ll 1$ and
$\alpha_W=g^2/(4\pi)\simeq 0.034$.
For $M=2$~TeV this is
$C_7^{\rm VLQ}\sim 10^{-8}$, negligible.

\emph{(iii)~$Z$-FCNC penguin.}
The tree-level $Z$-mediated $b\text{--}s$ FCNC coupling
(see below) generates a one-loop dipole through
a $Z$--$t$ penguin.
This carries an additional loop suppression
$\alpha_{\rm em}/(4\pi)\sim 6\times 10^{-4}$
beyond the tree-level $Z$-FCNC and is completely
negligible: $C_7^{Z\text{-FCNC}}\sim 10^{-5}$.

The chromomagnetic dipole $C_8$ has the same parametric
scaling as $C_7$ and is equally safe.
In summary, all one-loop dipole contributions satisfy
$|\delta C_7/C_7^{\rm SM}|\lesssim 10^{-2}$,
well below current and projected experimental
precision.

\subsection[Z-mediated rare decays and Rb]{$Z$-mediated rare decays and $R_b$}
\label{sec:Z-FCNC}

Integrating out the SU(2)-singlet VLQ chain states
at tree level generates a flavor-changing $Z$ coupling
\begin{equation}
\delta g_L^{bs}
= \frac{g}{2c_W}\;\frac{1}{2}\;\theta_L^b\,\theta_L^s,
\label{eq:dg-bs}
\end{equation}
where the factor $\tfrac{1}{2}=|I_3^{d_L}-I_3^{D}|$
reflects the isospin mismatch between the SM doublet
and the SU(2)-singlet VLQ.

\paragraph[Semileptonic operators C9, C10.]{Semileptonic operators $C_9$, $C_{10}$.}
Tree-level $Z$ exchange between the $b\text{--}s$
FCNC vertex and the lepton pair generates
contributions to the Wilson coefficients
$C_9$ (vector) and $C_{10}$ (axial):
\begin{align}
C_9^{\rm NP} &\sim \frac{\pi}{\alpha_{\rm em}}\,
\frac{\delta g_L^{bs}\,
(\!-\tfrac{1}{2}+2s_W^2)}{
\sqrt{2}\,G_F\,V_{tb}\,V_{ts}\,M_Z^2}\,,
\label{eq:C9-NP}\\
C_{10}^{\rm NP} &\sim -\frac{\pi}{\alpha_{\rm em}}\,
\frac{\delta g_L^{bs}\times\tfrac{1}{2}}{
\sqrt{2}\,G_F\,V_{tb}\,V_{ts}\,M_Z^2}\,.
\label{eq:C10-NP}
\end{align}
Because the $Z$-lepton vector coupling
$(-\tfrac{1}{2}+2s_W^2)\simeq -0.04$ is accidentally
small, $C_9^{\rm NP}$ is suppressed relative to
$C_{10}^{\rm NP}$ by a factor of
$\sim 2|{-}\tfrac{1}{2}+2s_W^2|\simeq 0.08$.
$C_{10}$ therefore provides the dominant constraint
from rare $B$ decays.

\paragraph[Bs to mu+mu-.]{$B_s\to\mu^+\mu^-$.}
The branching ratio scales as
$\mathrm{BR}\propto |C_{10}^{\rm SM}+C_{10}^{\rm NP}|^2$.
The current measurement
$\mathrm{BR}(B_s\to\mu^+\mu^-)
=(3.34\pm 0.27)\times 10^{-9}$~\cite{PDG2024}
is consistent with the SM prediction to
$\sim\!15\%$ precision, requiring
$|C_{10}^{\rm NP}|\lesssim 0.6$.

\paragraph[Rb vertex correction.]{$R_b$ vertex correction.}
The same SM--VLQ mixing that generates the FCNC also
shifts the diagonal $Zb\bar{b}$ coupling:
\begin{equation}
\delta g_L^{bb}
= \frac{1}{2}(\theta_L^b)^2,
\label{eq:dg-bb}
\end{equation}
producing a fractional change
$\delta R_b/R_b\simeq -2\,|g_L^b|\,\delta g_L^{bb}/
(g_L^{b\,2}+g_R^{b\,2})\simeq
-2.3\,(\theta_L^b)^2$.
The LEP measurement
$R_b=0.21629\pm 0.00066$~\cite{PDG2024}
constrains $|\delta R_b/R_b|\lesssim 0.3\%$,
yielding the bound
\begin{equation}
\theta_L^b\lesssim 0.04.
\label{eq:Rb-bound}
\end{equation}
This is the strongest individual constraint on
$\theta_L^b$ in the chain framework, tighter than the
oblique parameters.

\paragraph{Consistency.}
The mixing angle was estimated in
Eq.~\eqref{eq:tL-b} as
$\theta_L^b\sim\lambda_3\,v/M$, where $\lambda_3$ is
the $\mathcal{O}(1)$ endpoint coupling of the
third-generation doublet.
The $R_b$ bound requires
$\lambda_3\lesssim 0.04\,M/v\simeq 0.5$
for $M=2$~TeV, comfortably within the $\mathcal{O}(1)$
range.
For $\theta_L^b\sim 0.04$ the new-physics contributions
are:
\begin{equation}
\renewcommand{\arraystretch}{1.2}
\begin{array}{lrl}
\hline\hline
\text{Observable} & \text{NP} & \text{Bound} \\
\hline
|\delta C_7/C_7^{\rm SM}| & 2\times 10^{-3} & <0.10 \\
C_9^{\rm NP} & +0.02 & |C_9^{\rm NP}|<1 \\
C_{10}^{\rm NP} & -0.3 & |C_{10}^{\rm NP}|<0.6 \\
|\delta R_b/R_b| & 0.4\% & <0.6\% \\
C_{K\bar K} & 10^{-20}~\text{GeV}^{-2}
  & <10^{-12} \\
C_{B_s} & 10^{-16}~\text{GeV}^{-2}
  & <10^{-11} \\
\hline\hline
\end{array}
\label{eq:constraint-table}
\end{equation}
All constraints are satisfied, and the enormous
margins in the tree-level $\Delta F=2$ operators
confirm the efficacy of the lattice GIM mechanism.
The dominant constraints---$R_b$ and $C_{10}$---are
both sensitive to $(\theta_L^b)^2$ and thus to the
overall normalization of the third-generation
endpoint coupling, while lighter-generation processes
are additionally suppressed by powers of $\e$ from
the lattice hierarchy.

\section{Collider Phenomenology}
\label{sec:collider}

The common VLQ mass $M$ is the one free parameter of the
chain framework that is not fixed by the lattice
exponent structure.
Because the entry coupling scales as $v\,\e^{A_i/9}$
(from the Higgs VEV), the exit coupling scales as
$M\,\e^{B_j/9}$ (from the VLQ mass term), and the
chain propagator scales as
$(\kappa_1\kappa_2\kappa_3/M)\,\e^{\Sigma/9}$
[Eq.~\eqref{eq:Yd-pathsum}], the factor of $M$
cancels exactly in the effective Yukawa matrix.
All quark mass ratios, mixing angles, and CP phases
are therefore $M$-independent---they are determined
entirely by the lattice exponents and the
$\mathcal{O}(1)$ coefficients.

The chain mass enters the phenomenology only through
the SM--VLQ mixing angles
$\theta_L^i\sim\lambda_i\,v/M$
[Eq.~\eqref{eq:thetaL}], which control precision
observables, FCNC amplitudes, and collider production
rates.
From below, current ATLAS and CMS pair-production
searches exclude
$M\lesssim 1.6$~TeV~\cite{ATLAS:2024VLQ,CMS:2024VLQ}.
From above, there is no sharp theoretical bound:
higher $M$ simply reduces $\theta_L^i$ and
decouples the VLQ sector from low-energy observables,
while the flavor predictions remain intact.
However, for $M\gg v$ the endpoint couplings
$\lambda_i$ must be correspondingly larger to maintain
$\theta_L^b$ at a level consistent with the measured
$Zb\bar{b}$ vertex (Sec.~\ref{sec:Z-FCNC}); the
$R_b$ constraint $\theta_L^b\lesssim 0.04$ requires
$\lambda_3\lesssim 0.04\,M/v$, which remains
comfortably $\mathcal{O}(1)$ for
$M\lesssim 5$~TeV but begins to require
$\lambda_3> 1$ above that scale.
The benchmark range $M\sim 2$--$3$~TeV adopted
throughout this paper therefore represents the
window in which the framework is consistent with all
current constraints, the endpoint couplings are
naturally $\mathcal{O}(1)$, and the VLQ states are
accessible at the HL-LHC.

\subsection{Chain mass spectrum}

After the flavon develops its VEV, the four chain states
$D_1,\dots,D_4$ mix through the nearest-neighbor mass terms.
The physical masses are the eigenvalues of
$\mathcal{M}_D\mathcal{M}_D^\dagger$
[Eq.~\eqref{eq:MD-explicit}].
For degenerate bare masses $M_a=M$, the mass eigenvalues
of $\mathcal{M}_D\mathcal{M}_D^\dagger$ can be computed
exactly.
Since $\kappa_a\e^{h_a/9}$ is $\mathcal{O}(1)$
for the smallest hops (see Sec.~\ref{sec:convergence}),
the splittings are not parametrically small:
exact diagonalization gives physical masses
ranging from roughly $0.5\,M$ to $1.6\,M$ for $\kappa_a=1$.
For $M=2$~TeV the chain states thus span a range of
$\sim\!1$--$3$~TeV.
The lightest state ($D_4$, coupled most strongly to the
third generation) is produced first at colliders;
heavier chain states may decay via
$D_a\to D_{a+1}+\text{soft}$ if kinematically accessible,
producing cascade signatures with multiple $b$-jets
and $W/Z/H$ bosons.

\subsection{Pair production}

At the 14~TeV LHC, VLQ pair production proceeds via QCD and
is essentially model-independent.
Approximate cross sections are $\sim 10$, $1$, $0.1$, and $0.01$~fb
for $M_{\rm VLQ}=1.5$, $2.0$, $2.5$, and $3.0$~TeV,
respectively~\cite{Aguilar-Saavedra:2013qpa}.

\begin{figure*}[htbp]
\centering
\begin{tikzpicture}
\begin{axis}[
width=6.5cm,
height=4.5cm,
xlabel={$M_{\rm VLQ}$ (TeV)},
ylabel={Cross section (fb)},
ymode=log,
xmin=1.3, xmax=3.2,
ymin=0.005, ymax=30,
grid=major,
legend pos=north east,
legend style={font=\scriptsize}
]
\addplot[thick,blue,mark=*,mark size=1.5pt]
coordinates {(1.5,10) (2.0,1) (2.5,0.1) (3.0,0.01)};
\addlegendentry{Pair production}
\addplot[thick,red,dashed,mark=square*,mark size=1.5pt]
coordinates {(1.5,5) (2.0,2) (2.5,0.5) (3.0,0.1)};
\addlegendentry{Single ($\theta_L^b\sim 0.1$)}
\addplot[gray,thick,dotted] coordinates {(1.3,0.03) (3.2,0.03)};
\node[gray,font=\scriptsize] at (axis cs:2.5,0.06)
{HL-LHC $3\,\text{ab}^{-1}$ reach};
\end{axis}
\end{tikzpicture}
\caption{Approximate VLQ production cross sections at
$\sqrt{s}=14$~TeV for pair production (solid) and single
production with $\theta_L^b\sim 0.1$ (dashed).
The horizontal dotted line indicates the approximate
HL-LHC discovery threshold.}
\label{fig:VLQ-reach}
\end{figure*}
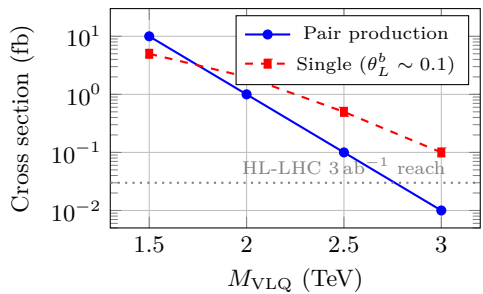

\noindent
Current ATLAS and CMS searches exclude VLQs with
masses below $1.5$--$1.8$~TeV (depending on the assumed
branching ratios), with the strongest limits reaching
$M>1.6$~TeV for singlet $B$-quarks decaying predominantly
via $bZ$ and $bH$~\cite{ATLAS:2024VLQ,CMS:2024VLQ}.
The chain framework with $M\sim 2$~TeV is therefore
consistent with all current limits, while remaining within
reach of the HL-LHC.
With $3~\text{ab}^{-1}$ at the HL-LHC, the discovery reach
extends to $M_{\rm VLQ}\approx 2.0$--$2.5$~TeV
(Fig.~\ref{fig:VLQ-reach}), depending
on decay channels and
systematics~\cite{Aguilar-Saavedra:2013qpa,ATLAS:2024VLQ}.
At a future 100~TeV $pp$ collider (FCC-hh), QCD pair
production would probe VLQ masses up to
$\sim\!10$~TeV~\cite{FCC:2018hh}, covering the entire
natural parameter space of the chain framework.

\subsection{Single production}

Single VLQ production scales as
\begin{equation}
\sigma_{\rm single} \propto (\theta_L^b)^2\,g_W^2,
\end{equation}
and is kinematically accessible to higher masses than pair
production.
For $\theta_L^b\sim 0.1$ the single-production cross section
can exceed pair production above $\sim 2$~TeV, making this
a sensitive probe of the chain coupling structure.

\subsection{Decay signatures}

The lightest chain state $D_4$ (which couples most strongly
to the third generation) decays as
\begin{equation}
D_4 \to bZ,\quad bH,\quad tW^-,
\end{equation}
with branching ratios approaching $1:1:2$ in the high-mass
limit (the Goldstone equivalence theorem).
Searches in the $tW$ and $bZ$ channels provide complementary
sensitivity~\cite{ATLAS:2024VLQ}.

\subsection{Electroweak precision constraints}

VLQs that mix with the third generation contribute to
the oblique parameters $S$ and $T$~\cite{Lavoura:1992np}.
For a single $SU(2)$-singlet VLQ of mass $M$ mixed with
$b_L$ at angle $\theta_L^b$, the leading contributions are
\begin{equation}
\Delta T \simeq \frac{3\,(\theta_L^b)^2}{16\pi s_W^2}
\frac{m_t^2}{M_W^2}\,\ln\frac{M^2}{m_t^2},
\end{equation}
\begin{equation}
\Delta S \simeq \frac{(\theta_L^b)^2}{6\pi}
\ln\frac{M^2}{m_t^2}.
\end{equation}
For $\theta_L^b\sim 0.1$ and $M=2$~TeV:
$\Delta T\simeq 0.01$ and $\Delta S\simeq 0.002$,
well within the current electroweak fit constraints
($\Delta T<0.1$, $\Delta S<0.1$ at $95\%$~CL)~\cite{PDG2024}.
The additional chain states $D_1$--$D_3$ have even smaller
mixing angles with SM quarks and contribute negligibly.

\section{Comparison with FN, Clockwork, and Chain Approaches}
\label{sec:comparison}

Table~\ref{tab:comparison} contrasts UF with the
standard Froggatt--Nielsen mechanism, clockwork/linear
dilaton models, and the recent TeV-chain framework of
AFHM~\cite{ArkaniHamed:2026chains}.

\begin{table*}[htbp]
\centering
\caption{Comparison of UF with standard FN,
clockwork, and the AFHM chain
framework~\cite{ArkaniHamed:2026chains}.}
\label{tab:comparison}
\setlength{\tabcolsep}{8pt}
\renewcommand{\arraystretch}{1.15}
\begin{tabular}{lcccc}
\toprule
Feature & FN & Clockwork & AFHM chains & UF \\
\midrule
Exponents & Free & Free & Free & Discrete \\
Discrete symmetry & No & No & No & $\Z_{18}$ \\
Algebraic closure & No & Partial & No & Yes \\
TeV states & No & Yes & Yes & Yes \\
FCNC protection & Param. & Geom. & Chain & Lattice \\
Quantitative fit & Coarse & Coarse & Good & Exact \\
Axion quality & No & No & No & Yes \\
Lepton extension & Ad hoc & Ad hoc & Yes & Yes \\
$\Bpar$ parameter & No & No & No & Yes \\
\bottomrule
\end{tabular}
\end{table*}

\noindent
The AFHM chain framework~\cite{ArkaniHamed:2026chains}
shares the core dynamical mechanism with UF:
nearest-neighbor VLQ chains at the TeV scale generate
hierarchical Yukawa matrices, and chain locality provides
FCNC suppression.
However, UF is more complete in several respects:

\emph{(i)~Discrete exponent quantization.}
The AFHM framework treats the chain couplings as free
continuous parameters, allowing any hierarchy to be
generated by suitable tuning.
UF, by contrast, enforces a $\Z_{18}$
discrete gauge symmetry that quantizes all Yukawa
exponents on a ninths lattice.
This eliminates continuous parametric freedom in the
exponents and makes the hierarchy a structural consequence
of the symmetry, not a choice of parameters.

\emph{(ii)~Exact $\mathcal{O}(1)$ coefficients.}
With continuous coupling freedom, the AFHM framework
necessarily fits the mass spectrum by adjusting magnitudes.
The $\Bpar$-lattice textures of
Eq.~\eqref{eq:pud-Blattice} reproduce all six quark
masses with $\mathcal{O}(1)$ coefficients that are
essentially unity ($c_f\simeq 1.00$), with no
parametric freedom remaining.

\emph{(iii)~Axion quality protection.}
The $\Z_{18}$ symmetry that enforces the flavor
lattice simultaneously protects the Peccei--Quinn symmetry
against gravitational spoiling up to the required operator
dimension~\cite{FlavorInNinths}.
This connection between the flavor hierarchy and the
strong CP problem is absent in the AFHM framework, where
the chain couplings carry no discrete gauge charge.

\emph{(iv)~Algebraic closure.}
The ninths lattice is algebraically closed under the
operations that appear in the Yukawa
construction---addition of exponents (chain hops),
multiplication by the expansion parameter (flavon
insertions), and the seesaw-like corrections from
diagonalization.
This closure is a structural consequence of the
$\Z_9$ discrete gauge symmetry underlying the
specific value $\Bpar=75/14$, and has no counterpart
in a continuous-parameter framework.

\emph{(v)~Single-parameter organization.}
The entire quark and lepton mass spectrum is organized by
the single parameter $\Bpar=75/14$
($\e=1/\Bpar\simeq 0.187$).
The AFHM framework, while economical in its field content,
requires independent specification of each chain coupling
constant.

\section{Lepton Sector Extension}
\label{sec:leptons}

The $\Bpar$-lattice framework extends naturally to the
lepton sector using the same expansion parameter
$\e=1/\Bpar$ and the same discrete gauge
structure~\cite{Barger2025bfnb,LeptonLattice}.
The construction rests on two ingredients:
(i)~$B$-lattice power counting, which fixes the
charged-lepton Yukawa suppressions and neutrino mass
eigenvalue ratios, and
(ii)~an approximate $\mu$--$\tau$ symmetry, which fixes
the $\mathcal{O}(1)$ structure of the neutrino $23$ block.

\subsection{Charged-lepton texture}

The charged-lepton exponent matrix, expressed on the
denominator-$18$ lattice, is~\cite{LeptonLattice}
\begin{equation}
p^e=\frac{1}{18}
\begin{pmatrix}
87&64&40\\
64&30&36\\
40&36&0
\end{pmatrix},
\qquad
(\Ye)_{ij}=c^e_{ij}\,\e^{\,p^e_{ij}}.
\label{eq:pe-texture}
\end{equation}
This symmetric texture departs from strict charge
additivity in the $(2,3)$ sector, where the off-diagonal
suppression $\e^2$ is stronger than the additive rule
would predict; this additional suppression ensures that
the charged-lepton contribution to atmospheric mixing
remains a perturbative correction to the
neutrino-dominant structure.

The diagonal entries fix the charged-lepton mass hierarchy:
\begin{equation}
m_e : m_\mu : m_\tau
\sim c_e\,\e^{29/6} : c_\mu\,\e^{5/3} : 1,
\label{eq:me-hierarchy}
\end{equation}
with $c_e\simeq 0.93$ and $c_\mu\simeq 0.97$---both
essentially unity.

The left-handed diagonalization matrix is
\begin{equation}
\Ue \simeq
\begin{pmatrix}
1 & \e^{17/9} & \e^{20/9}\\
-\e^{17/9} & 1 & \e^2\\
-\e^{20/9} & -\e^2 & 1
\end{pmatrix},
\label{eq:UeL}
\end{equation}
so the charged-lepton mixing angles are
$\theta_{12}^e\sim\e^{17/9}\simeq 0.04$,
$\theta_{23}^e\sim\e^2\simeq 0.03$,
$\theta_{13}^e\sim\e^{20/9}\simeq 0.03$.

In the chain picture, the charged-lepton sector is
realized by an analogous vectorlike lepton chain
$E_a + \bar{E}_a$, with the same hop structure
$\{1,2,4\}/9$ as the quark chains.

\subsection[Neutrino mass matrix and mu--tau symmetry]{Neutrino mass matrix and $\mu$--$\tau$ symmetry}

Light neutrino masses are generated through the
Weinberg operator~\cite{Barger2025bfnb}.
The $B$-lattice hierarchy alone does not determine
the neutrino mixing pattern; large atmospheric and solar
angles require additional $\mathcal{O}(1)$ structure.
The FN charge assignment
$p^\nu_{L_1}\gg p^\nu_{L_2}=p^\nu_{L_3}$
produces the exponent texture~\cite{LeptonLattice}
\begin{equation}
p^\nu=\frac{1}{9}
\begin{pmatrix}
9 & 9 & 9\\
9 & 0 & 0\\
9 & 0 & 0
\end{pmatrix},
\label{eq:pnu-texture}
\end{equation}
which suppresses first-generation couplings by
$\mathcal{O}(\e)$ relative to the $23$ block.
The $\mathcal{O}(1)$ coefficients in the $23$ block
must satisfy approximate $\mu$--$\tau$
symmetry~\cite{HarrisonScott,Lam,MuTauReview}:
$c^\nu_{22}\simeq c^\nu_{33}$ and
$c^\nu_{12}\simeq c^\nu_{13}$, together with a
near-maximal ratio $c^\nu_{23}/c^\nu_{22}\simeq -1$.
This structure can be enforced by a
$Z_2^{\mu\tau}$ exchange symmetry or by larger discrete
groups such as $A_4$ or
$S_4$~\cite{King:2013eh,MaRajasekaran}.

The resulting leading-order neutrino mass matrix is
\begin{equation}
\frac{m_\nu}{m_3}\;\simeq\;
\begin{pmatrix}
\mathcal{O}(\e) & \mathcal{O}(\e) & \mathcal{O}(\e)\\
\mathcal{O}(\e) & \;\;\;\tfrac12 & -\tfrac12 \\
\mathcal{O}(\e) & -\tfrac12 & \;\;\;\tfrac12
\end{pmatrix},
\end{equation}
whose $23$ block has eigenvalues $0$ and $1$ (in units
of $m_3$), giving maximal atmospheric mixing and a
near-tribimaximal solar angle.
The neutrino mass hierarchy in normal ordering is
$m_3:m_2:m_1\sim 1:\e:\e^2$,
with $m_3\simeq 50$~meV, $m_2\simeq 9$~meV,
$m_1\simeq 2$~meV, and $\sum m_i\simeq 61$~meV
(compatible with cosmological bounds).

The neutrino diagonalization matrix is near-tribimaximal:
\begin{equation}
\Unu\simeq
\begin{pmatrix}
\frac{2}{\sqrt6}&\frac{1}{\sqrt3}&\mathcal{O}(\e)\\[4pt]
-\frac{1}{\sqrt6}&\frac{1}{\sqrt3}&\frac{1}{\sqrt2}\\[4pt]
-\frac{1}{\sqrt6}&\frac{1}{\sqrt3}&-\frac{1}{\sqrt2}
\end{pmatrix},
\label{eq:Unu}
\end{equation}
with $\theta_{13}^\nu\sim\e$ generated by
$\mathcal{O}(\e)$ breaking of the $Z_2^{\mu\tau}$.

\subsection[PMNS mixing: octant--delta theorem]{PMNS mixing: octant--$\delta$ theorem}

The PMNS matrix is
$U_{\rm PMNS} = U_{eL}^\dagger\,U_\nu$.
Since the charged-lepton angles are small
($\sim\!\e^{17/9}$--$\e^2$), the large PMNS angles
are neutrino-dominant, with $U_{eL}$ providing
perturbative corrections.

The reactor element receives contributions from both
sectors~\cite{LeptonLattice}:
\begin{equation}
U_{e3} \simeq s_{13}^\nu e^{-i\delta_\nu}
- \theta_{12}^e\,s_{23}^\nu\,e^{i\phi_{12}^e}
- \theta_{13}^e\,c_{23}^\nu\,e^{i\phi_{13}^e}.
\end{equation}
Defining $r\equiv\theta_{12}^e s_{23}^\nu/s_{13}^\nu$
and $\Phi_e\equiv\phi_{12}^e+\delta_\nu$,
the observed Dirac phase satisfies the
\emph{octant--$\delta$ theorem}:
\begin{equation}
\boxed{
\delta \simeq \delta_\nu
- \arg\!\left(1 - r\,e^{-i\Phi_e}\right),
}
\label{eq:delta-theorem}
\end{equation}
and the atmospheric angle receives its dominant
correction from the charged-lepton $23$ rotation:
\begin{equation}
\boxed{
\theta_{23} \simeq \theta_{23}^\nu
- \theta_{23}^e \cos\phi_{23}^e.
}
\label{eq:theta23-theorem}
\end{equation}

The single-flavon origin of the $B$-lattice textures
ensures~\cite{LeptonLattice} that
$\phi_{12}^e\approx\phi_{23}^e$ to within
$\sim\!\phi_0/9$, where $\phi_0=\arg\langle\Phi\rangle$
is the flavon phase.
Under this alignment, the sign of the octant shift
$(\theta_{23}-45^\circ)$ correlates directly with the
direction of the $\delta$ shift, producing the
two-branch prediction.

\subsection{Two-branch prediction}
\label{sec:two-branch}

The interplay between Eqs.~\eqref{eq:delta-theorem}
and~\eqref{eq:theta23-theorem} yields two
normal-ordering solutions~\cite{LeptonLattice}:

\begin{table*}[htbp]
\centering
\caption{Two-branch prediction for the atmospheric
octant and Dirac CP phase.  The Jarlskog invariant
and effective Majorana mass are
branch-independent.}
\label{tab:NO-branches}
\setlength{\tabcolsep}{4pt}
\renewcommand{\arraystretch}{1.1}
\begin{tabular}{lccc}
\toprule
Observable & NO-I & NO-II & Discrim.?\\
\midrule
$\theta_{23}$ & $43^\circ$ & $46^\circ$ & $\checkmark$ \\
$\delta$ & $286^\circ$ & $304^\circ$ & $\checkmark$ \\
$J_{\rm CP}$ & $-0.027$ & $-0.025$ & $\times$ \\
$m_{\beta\beta}$ [meV] & $3.3$ & $3.6$ & $\times$ \\
\bottomrule
\end{tabular}
\end{table*}

\noindent
NO-I (lower octant) is favored $\sim\!4{:}1$ by a
theoretical prior based on the coefficient scan.
The framework makes a falsifiable prediction: if future
data establish the upper octant, $\delta$ must lie near
$304^\circ$; if the lower octant, near $286^\circ$.
Discovery of $\delta$ in the first or second quadrant
($0^\circ$--$180^\circ$) in either octant would rule
out the single-$\Bpar$ lattice textures with aligned
phases.
The decisive measurements are $\nu_\mu\to\nu_e$
appearance rates at DUNE and Hyper-Kamiokande, combined
with octant determination from atmospheric neutrinos
at IceCube and independent mass-ordering resolution
by JUNO.

Three observables are nearly branch-independent and
cannot discriminate:
$J_{\rm CP}\simeq -0.027$,
$m_{\beta\beta}\simeq 3$--$4~\text{meV}$, and
$\sin^2\theta_{13}\simeq 0.022$.

\subsection{Neutrinoless double beta decay}

In normal ordering with
$m_3:m_2:m_1\sim 1:\e:\e^2$, the effective Majorana mass
scales as
\begin{equation}
m_{\beta\beta}\sim \e\,m_3
\sim \frac{m_3}{\Bpar}
\simeq 9~\text{meV},
\end{equation}
within the projected reach of next-generation
$0\nu\beta\beta$ experiments.
The coefficient scan gives a tighter range
$m_{\beta\beta}\simeq 3$--$4$~meV for vanishing
Majorana phases (Table~\ref{tab:NO-branches}).

\section{Quark--Lepton Connections}
\label{sec:ql-connections}

The $\Bpar$-lattice framework reveals a set of
quark--lepton connections that go beyond the usual
unification paradigm.
In this section we show that the charged-lepton geometric
mean ratio (GMR) is directly determined by the lattice
parameter $\e$, and that this identification provides a
single organizing principle for both CKM and PMNS
mixing magnitudes.

\subsection[The charged-lepton GMR and epsilon 3/2]{The charged-lepton GMR and $\e^{3/2}$}

Define the charged-lepton geometric mean ratio
\begin{equation}
\beta \;\equiv\;
\frac{\sqrt{m_e^2\,m_\tau^2}}{m_\mu^2}
\;=\; \frac{m_e\,m_\tau}{m_\mu^2}.
\label{eq:beta-def}
\end{equation}
Using $\overline{\rm MS}$ masses at $M_Z$, one finds
$\beta=0.0806\pm 0.0001$.
In the $\Bpar$-lattice, the charged-lepton mass scaling
$m_e:m_\mu:m_\tau\sim\e^{p_e}:\e^{p_\mu}:1$ gives
\begin{equation}
\beta
= \e^{p_e-2p_\mu}
= \e^{29/6 - 2\times 5/3}
= \e^{29/6 - 10/3}
= \e^{3/2},
\end{equation}
and numerically $\e^{3/2}\simeq 0.0806$---an exact match.
The exponent $3/2 = Q(L_1)+Q(L_2)=1+\tfrac{1}{2}$ is
the sum of the first- and second-generation
lepton-doublet charges, so $\beta$ is \emph{determined by
the $\Z_{18}$ discrete gauge charges}.

Equivalently,
\begin{equation}
\frac{1}{\beta}=\Bpar^{3/2}\simeq 12.40,
\label{eq:beta-B}
\end{equation}
providing a direct algebraic link between the
charged-lepton mass spectrum and the single
$\Bpar$-lattice organizing parameter.

Geometric mean mass ratios arise as the natural
organizing variables whenever the mass spectrum is a
power law of a single parameter: if
$m_i\propto\e^{p_i}$ then
$\sqrt{m_i m_j}/m_k=\e^{(p_i+p_j-2p_k)/2}$ is
itself a power of $\e$, so all such ratios lie on
the lattice.
This structure is present in any
Froggatt--Nielsen-type model with a single
flavon~\cite{FroggattNielsen1980} and is realized
concretely in the UF framework through the
multi-messenger sum
(Sec.~\ref{sec:Blattice-connection}) that generates
the $\mathcal{O}(1)$ coefficients $C^f_{ij}$.

\subsection[CKM magnitudes as powers of beta]{CKM magnitudes as powers of $\beta$}

The $\Bpar$-lattice CKM scaling can be reexpressed
entirely in terms of $\beta$.  Since $\e=\beta^{2/3}$:
\begin{align}
|V_{us}|&\sim\e^{8/9}
 =\beta^{16/27},
\label{eq:Vus-beta}\\
|V_{cb}|&\sim\e^{17/9}
 =\beta^{34/27},
\label{eq:Vcb-beta}\\
|V_{ub}|&\sim\e^{10/3}
 =\beta^{20/9}.
\label{eq:Vub-beta}
\end{align}
The CKM hierarchy is thus controlled by a single
number $\beta$ that is itself a ratio of charged-lepton
masses.
Two cross-checks are immediate:
\begin{align}
\frac{|V_{cb}|}{|V_{us}|}
&=\beta^{34/27-16/27}=\beta^{2/3},
\label{eq:Vcb-Vus-ratio}\\
\frac{|V_{ub}|}{|V_{us}|}
&=\beta^{20/9-16/27}=\beta^{44/27}.
\label{eq:Vub-Vus-ratio}
\end{align}

\subsection[PMNS magnitudes as powers of beta]{PMNS magnitudes as powers of $\beta$}

The PMNS mixing angles also organize as powers
of $\beta$.
A Froggatt--Nielsen-type parameterization with
denominator~$25$ exponents~\cite{LeptonLattice} gives
\begin{align}
\sin\theta_{12}&\simeq\beta^{6/25}\simeq 0.547,
\label{eq:s12-beta}\\
\sin\theta_{13}&\simeq\beta^{19/25}\simeq 0.148,
\label{eq:s13-beta}\\
\sin\theta_{23}&\simeq\beta^{4/25}\simeq 0.668,
\label{eq:s23-beta}
\end{align}
corresponding to
$\theta_{12}\simeq 33^\circ$,
$\theta_{13}\simeq 8.5^\circ$,
$\theta_{23}\simeq 42^\circ$ (lower octant).
The Dirac CP phase is
\begin{equation}
\delta\simeq(1/\beta)^{14/25}\simeq 235^\circ.
\end{equation}
These values are consistent with the NO-I solution
of the lepton lattice paper
(Sec.~\ref{sec:two-branch}).

Three internal relations follow:
\begin{itemize}
\item
$\theta_{12}^{\rm PMNS}+\theta_{12}^{\rm CKM}
\simeq 33^\circ+13^\circ\simeq 46^\circ$ (QLC) and\\[6pt]
$\theta_{23}^{\rm PMNS}+\theta_{23}^{\rm CKM}
\simeq 42^\circ+2.4^\circ\simeq 44^\circ$ (QLC),\\[6pt]
both close to $45^\circ$;
\item
$\sin\theta_{12}\times\sin\theta_{13}
=\beta^{6/25+19/25}=\beta$ (QLC);
\item
$\sin\theta_{23}=(\sin\theta_{12})^{2/3}$, since
$4/25=(2/3)\times 6/25$.
\end{itemize}

\noindent
Here QLC denotes quark--lepton
complementarity~\cite{Raidal:2004iw,MinakataSmirnov}.

The full PMNS magnitude matrix has the FN representation
\begin{equation}
|U_{\rm PMNS}| \;\approx\;
\begin{pmatrix}
\beta^{2/25} & \beta^{6/25} & \beta^{19/25}\\[3pt]
\beta^{10/25}& \beta^{4/25} & \beta^{4/25}\\[3pt]
\beta^{8/25} & \beta^{13/50}& \beta^{3/25}
\end{pmatrix}.
\label{eq:PMNS-FN}
\end{equation}
Since $\beta=\e^{3/2}$, each exponent $p/25$
translates to an $\e$-exponent $3p/50$,
connecting the PMNS structure back to the
$\Bpar$-lattice.

Thus both CKM and PMNS mixing hierarchies are
controlled by rational powers of the same
charged-lepton GMR $\beta=\e^{3/2}$.

\paragraph{Relation to the near-tribimaximal structure.}
The $\beta$-power parameterization and the
near-tribimaximal (TBM) construction of
Sec.~\ref{sec:leptons} are complementary descriptions
of the same physics.
Exact TBM predicts
$\sin\theta_{12}=1/\sqrt{3}\simeq 0.577$,
$\sin\theta_{13}=0$, and
$\sin\theta_{23}=1/\sqrt{2}\simeq 0.707$.
The $\beta$-power values
$\sin\theta_{12}\simeq 0.547$,
$\sin\theta_{13}\simeq 0.148$,
$\sin\theta_{23}\simeq 0.668$
depart from TBM by $\sim\!5\%$---precisely the
$\mathcal{O}(\e)$ magnitude of the charged-lepton
corrections $\theta^e_{ij}\sim\e^{17/9}$--$\e^2$
derived in Sec.~\ref{sec:leptons}.
Thus the near-TBM framework provides the
\emph{structural origin} of the PMNS pattern
(approximate $\mu$--$\tau$ symmetry in $U_\nu$,
broken by $U_{eL}$), while the $\beta$-power
representation captures the \emph{algebraic result}
after the charged-lepton corrections have acted.
Numerically, the two parameterizations agree at the
sub-percent level on all three mixing angles:

\begin{center}
\setlength{\tabcolsep}{5pt}
\renewcommand{\arraystretch}{1.05}
\begin{tabular}{lccccc}
\toprule
& TBM & Near-TBM & $\beta^p$ & NuFIT \\
& (exact) & (NO-I) & & (NO) \\
\midrule
$\sin\theta_{12}$ & 0.577 & 0.545 & 0.547 & 0.550\\
$\sin\theta_{13}$ & 0 & 0.148 & 0.148 & 0.148\\
$\sin\theta_{23}$ & 0.707 & 0.682 & 0.668 & 0.672\\
$\theta_{23}$ octant & max & lower & lower & lower\\
\bottomrule
\end{tabular}
\end{center}

\noindent
Both select the NO-I (lower-octant) solution and are
consistent with NuFIT~6.0.

\subsection{Dual mass representations of CKM elements}

Because every quark mass is a definite power of
$\e=1/\Bpar$, each CKM magnitude can be expressed
equivalently in terms of down-type mass ratios
($d,s,b$), up-type mass ratios ($u,c,t$), or mixed
combinations.
We highlight several notable consequences; a
systematic treatment is given elsewhere.

\paragraph{Dual representations.}
The Cabibbo element admits four equivalent forms:
\begin{equation}
|V_{us}|
=\Bigl(\frac{d}{s}\Bigr)^{\!1/2}
=\frac{c}{b}
=\Bigl(\frac{u}{t}\Bigr)^{\!1/8}
=\Bigl[\frac{(ut)^{1/2}}{c}\Bigr]^{4}.
\label{eq:Vus-dual}
\end{equation}
Each form equals $\e^{8/9}\simeq 0.225$.
Similarly,
\begin{equation}
|V_{cb}|=\Bigl(\frac{u}{c}\Bigr)^{\!1/2}
=\frac{db}{ut},
\qquad
|V_{ub}|=\frac{c}{t}=\frac{ds}{uc}.
\end{equation}

\paragraph{Mass sum rules.}
Equating dual representations yields algebraic
relations among quark masses.
Two illustrative ones are
\begin{equation}
uc^2 = dsb,
\qquad
\frac{(ub)^{1/2}}{s}=1,
\label{eq:sum-rules}
\end{equation}
both satisfied to sub-permille accuracy by the
$\Bpar$-lattice benchmark masses.
The first can be written as
\begin{equation}
\frac{tcu}{bsd}
=\frac{1}{V_{ub}}
=\Bpar^{10/3}\simeq 269,
\label{eq:tcu-bsd}
\end{equation}
expressing the full quark mass product ratio
as a power of $\Bpar$.
Also note that
\begin{equation}
\Bpar = \left(\frac{tc}{sd}\right)^{\!1/8}.
\label{eq:B-from-tcsd}
\end{equation}

These identities are not numerological coincidences:
they are \emph{algebraic consequences} of the fact
that all six quark masses are determined by a single
parameter $\Bpar$ and the discrete $\Z_9$ charges.
In the $\Bpar$-lattice, the masses scale as
$(u,c,t)=t\,(\e^{64/9},\,\e^{10/3},\,1)$ and
$(d,s,b)=b\,(\e^{37/9},\,\e^{7/3},\,1)$,
so any ratio of mass monomials reduces to a
rational power of $\e$.
A complete catalog of the resulting dual CKM
representations and geometric mean relations
will be presented in a forthcoming companion paper.

\subsection{Color-factor mass relation and the golden ratio}

Two additional quark--lepton mass relations follow from
the $\Bpar$-lattice charge assignments:
\begin{enumerate}
\item \emph{Color factor.}
The geometric means of the down-quark and charged-lepton
mass-squares satisfy
\begin{equation}
\bigl(d^2 s^2 b^2\bigr)^{1/3}
= 3\,\bigl(e^2\mu^2\tau^2\bigr)^{1/3},
\end{equation}
to sub-percent accuracy.
The factor of~3 suggests an underlying SU(3)$_c$
color-weight connection between the two sectors.

\item \emph{Golden ratio.}
The $b$-quark to $\tau$-lepton mass ratio is
\begin{equation}
\frac{m_b}{m_\tau}
= \frac{m_b^{\rm ref}}{m_\tau^{\overline{\rm MS}}(M_Z)}
\simeq \varphi \equiv \frac{1+\sqrt{5}}{2}
\simeq 1.618,
\label{eq:golden}
\end{equation}
as previously noted in
Paper~I~\cite{Barger2025bfn}.
This relation is used to set the reference scale
$m_b^{\rm ref}=\varphi\,m_\tau^{\overline{\rm MS}}(M_Z)$
in the quark-mass benchmark of
Sec.~\ref{sec:Blattice-connection}.
\end{enumerate}

These relations, together with the $\beta$
parameterization of CKM and PMNS magnitudes,
provide evidence that the quark and lepton mass spectra
are organized by a common principle rooted in the
$\Bpar$-lattice charge assignments.

\section{Discrete Anomaly Cancellation}
\label{sec:anomaly}

A discrete gauge symmetry is physical---and in particular
can protect the axion quality---only if it is free of
gauge anomalies.
For a $Z_N$ symmetry, the Ib\'a\~nez--Ross
conditions~\cite{IbanezRoss} are the discrete analogs
of continuous anomaly cancellation.
Writing $q_i$ for the $Z_N$ charge of each left-handed
Weyl fermion, the conditions require
\begin{align}
A_3 &\equiv \sum_{\rm color}\! q_i\,d_2(i)
\equiv 0 \pmod{N}, \label{eq:A3}\\
A_2 &\equiv \sum_{\rm weak}\! q_i\,d_3(i)
\equiv 0 \pmod{N}, \label{eq:A2}\\
A_{\rm grav} &\equiv \sum_{\rm all}\! q_i\,d_3(i)\,d_2(i)
\equiv 0 \pmod{N}, \label{eq:Agrav}
\end{align}
where $d_2$ and $d_3$ denote the SU(2)$_L$ and SU(3)$_c$
representation dimensions, and the sums run over the
indicated subsets of fermions.

\subsection[Z18 charge assignments]{$\Z_{18}$ charge assignments}

The complete $\Z_{18}$ charge table follows from the
$\Bpar$-lattice FN charges determined in
Secs.~\ref{sec:Blattice-connection}
and~\ref{sec:leptons}~\cite{Barger2025bfnb,FlavorInNinths},
expressed as integer charges $q_{18}\equiv 18\,Q$:
\begin{equation}
\renewcommand{\arraystretch}{1.2}
\begin{array}{lcc}
\hline\hline
\text{Field} & Q(\cdot) & q_{18}\\
\hline
Q_{1,2,3} & 3,\;2,\;0 & 54,\;36,\;0\\
u^c_{1,2,3} & \tfrac{37}{9},\;\tfrac{4}{3},\;0
 & 74,\;24,\;0\\
d^c_{1,2,3} & \tfrac{10}{9},\;\tfrac{1}{3},\;0
 & 20,\;6,\;0\\
L_{1,2,3} & 1,\;\tfrac{1}{2},\;0 & 18,\;9,\;0\\
e^c_{1,2,3} & \tfrac{10}{3},\;\tfrac{1}{6},\;0
 & 60,\;3,\;0\\
N_{1,2,3} & 0,\;0,\;0 & 0,\;0,\;0\\
\Phi & \tfrac{1}{18} & 1\\
H & 0 & 0\\
\hline\hline
\end{array}
\label{eq:Z18-charges}
\end{equation}
Here $N_i$ are right-handed (sterile) neutrinos and
$H$ is the SM Higgs doublet.
The third generation of all SM fields carries zero
charge, so the anomaly sums receive contributions only
from the first two generations.
The $\Z_9$ subgroup (denominator-$9$ lattice) controls
quark-sector exponents, while the full $\Z_{18}$
is required to accommodate the half-integer
lepton-doublet charge $Q(L_2)=\tfrac{1}{2}$.

\paragraph{Uniqueness of the charge table.}
It is important to emphasize that the charges in
Eq.~\eqref{eq:Z18-charges} are not one of many
possible anomaly-free solutions selected to match the
data.
The logical sequence is reversed: the quark-sector
charges $Q(Q_i)$ and $Q(f^c_j)$ are uniquely fixed by
the observed mass ratios and the value
$\Bpar=75/14$ (Sec.~\ref{sec:Blattice-connection}),
the lepton-sector charges are similarly determined by
the charged-lepton and neutrino mass hierarchies
(Sec.~\ref{sec:leptons}), and the anomaly conditions
are then a nontrivial \emph{consistency check} on these
data-determined charges.
Because the charges are overdetermined by the mass
spectrum, one cannot freely adjust them to satisfy the
anomaly conditions---the fact that they do so
is a structural property of the $\Bpar$-lattice, not
a consequence of parameter tuning.

\emph{Bottom-up origin and uniqueness.}
The charges in Eq.~\eqref{eq:Z18-charges} are not
free parameters adjusted to satisfy anomaly
cancellation.
They are determined bottom-up by the observed
fermion mass spectrum through the $\Bpar$-lattice
formula $m_f=m_b^{\rm ref}\cdot\e^{n_f}$
(Paper~I~\cite{Barger2025bfn}): the diagonal
exponents $n_f$ are read off from the measured quark
and lepton mass ratios, and the
factorization $p^f_{ij}=Q(Q_i)+Q(f^c_j)$ then
determines all charges up to the convention $Q(\cdot_3)=0$
for the third generation.
Within the $\Bpar$-lattice framework, these charges
are therefore unique---they are the only set that
simultaneously reproduces the six quark mass ratios,
three CKM magnitudes, and three charged-lepton masses
with $\mathcal{O}(1)$ coefficients $c_f\simeq 1.00$
and a single expansion parameter $\e=1/\Bpar$.
That the resulting $q_{18}$ values happen to satisfy
the Ib\'a\~nez--Ross anomaly
conditions~\eqref{eq:A3}--\eqref{eq:Agrav}
(modulo the Green--Schwarz shift of
Sec.~\ref{sec:GS})
is a nontrivial \emph{consistency check}---a
prediction of the framework, not an input.

\subsection{Neutrino masses from the charge table}

The $\Z_{18}$ charges determine the neutrino mass
texture through two complementary mechanisms.

\paragraph{Weinberg operator.}
The dimension-five operator
$L_i L_j H H/\Lambda$ carries $\Z_{18}$ charge
$Q(L_i)+Q(L_j)$, so the required number of flavon
insertions gives the exponent matrix
$p^\nu_{ij}=Q(L_i)+Q(L_j)$ of
Eq.~\eqref{eq:pnu-texture}.

\paragraph{Type-I seesaw.}
With $Q(N_i)=0$, the Dirac Yukawa matrix scales as
$(Y_\nu)_{ij}\sim\e^{Q(L_i)}$ (independent of the
right-handed generation index), and the Majorana mass
matrix $M_N$ is unsuppressed.
Integrating out $N$ via the seesaw formula
$(m_\nu)_{ij}\sim v^2(Y_\nu M_N^{-1}Y_\nu^T)_{ij}$
reproduces
\begin{equation}
(m_\nu)_{ij}
\sim \frac{v^2}{\Lambda_N}\,\e^{Q(L_i)+Q(L_j)},
\end{equation}
recovering the Weinberg-operator texture with the
identification $\Lambda_N\sim\Lambda$.
The diagonal entries give the mass eigenvalue hierarchy
\begin{equation}
m_1:m_2:m_3\sim\e^{2Q(L_1)}:\e^{2Q(L_2)}:\e^{2Q(L_3)}
=\e^2:\e:1,
\end{equation}
in agreement with normal ordering.

The vanishing $\Z_{18}$ charge of $N_i$ has two
further consequences: the right-handed neutrinos
are SM singlets with zero discrete charge, so they
do not contribute to any of the anomaly
sums~\eqref{eq:A3}--\eqref{eq:Agrav}; and their
Majorana mass $M_N\sim\Lambda$ is unsuppressed by the
flavon, placing the seesaw scale at
$\Lambda\sim 5\times 10^{11}$~GeV, consistent with
the axion decay constant $f_a$ discussed in
Sec.~\ref{sec:axion}.

\subsection{Green--Schwarz cancellation and the axion}
\label{sec:GS}

The SM fermion charges above do not by themselves
satisfy Eqs.~\eqref{eq:A3}--\eqref{eq:Agrav}; the
residual anomalies $A_a\not\equiv 0\pmod{18}$ are
nonzero.
This is a generic feature of FN models and is resolved
by the Green--Schwarz (GS) mechanism: the $\Z_{18}$
discrete gauge symmetry descends from a continuous
$U(1)_{\rm FN}$ that is spontaneously broken by the
flavon VEV $\langle\Phi\rangle$.
The $U(1)_{\rm FN}$ mixed anomalies are cancelled by
a GS axion, and this cancellation persists in the
discrete $Z_N$ remnant~\cite{IbanezRoss,BanksDine}.

The connection to the strong CP problem is immediate:
the same PQ axion that solves the strong CP problem
(Sec.~\ref{sec:axion}) can serve as the GS field that
cancels the discrete gauge anomalies.
This identification is natural in the ``Flavor in
Ninths'' framework~\cite{FlavorInNinths}, where the
$\Z_{18}$ discrete gauge symmetry simultaneously
\begin{enumerate}
\item enforces the ninths-quantized lattice exponents,
\item protects the PQ symmetry from Planck-suppressed
operators, and
\item is rendered anomaly-free by the PQ/GS axion.
\end{enumerate}
The VLQ chain fields ($D_a+\bar{D}_a$,
$U_a+\bar{U}_a$) are vectorlike under all gauge groups
including $\Z_{18}$ and therefore do not contribute to
the anomaly sums.

\subsection{Spectator fermion alternative}

If the GS mechanism is not invoked, anomaly cancellation
can alternatively be achieved by introducing spectator
fermions at a high scale $\Lambda\gg M_{\rm VLQ}$ that
carry appropriate $\Z_{18}$ charges but are SM singlets.
Such spectators are common in UV completions of FN
models and decouple from low-energy physics.
In either case, the anomaly-free $\Z_{18}$ is a
legitimate discrete gauge symmetry, and its protection
of the axion quality and the lattice exponent structure
is on firm theoretical footing.

\section{Connection to the Axion Quality Problem}
\label{sec:axion}

The ``Flavor in Ninths'' paper~\cite{FlavorInNinths} demonstrated
that the same discrete $Z_N$ symmetry enforcing the ninths lattice
can simultaneously address the Peccei--Quinn (PQ) axion quality
problem~\cite{PecceiQuinn}.
The axion quality problem---the requirement that
Planck-suppressed operators violating the global
$U(1)_{\rm PQ}$ symmetry be forbidden to sufficiently
high dimension---was identified by
Kamionkowski and March-Russell~\cite{Kamionkowski:1992mf},
Holman \textit{et al.}~\cite{Holman:1992us},
and Barr and Seckel~\cite{Barr:1992qq}, and
has been studied quantitatively in the context of the
string landscape by
Baer, Barger, and Sengupta~\cite{Baer:2018avn}
and in gauge-protected axion models by
Bhattiprolu and Martin~\cite{Bhattiprolu:2021sli}.
In the UF construction, the FN flavon $\Phi$ is identified with
the PQ scalar, and the $\Z_{18}$ discrete gauge symmetry forbids
all PQ-violating Planck-suppressed operators below dimension~$9$.

The idea of connecting the flavor symmetry to axion quality
has been pursued independently by
Babu, Chandrasekar, and
Tavartkiladze~\cite{Babu:2026axion}, who use a gauged
continuous $U(1)_F$ symmetry to realize the FN mechanism
while an accidental global $U(1)_{\rm PQ}$ emerges as a
byproduct, yielding a high-quality
Dine--Fischler--Srednicki--Zhitnitsky
(DFSZ)-type axion~\cite{Dine:1981rt,Zhitnitsky:1980tq}.
UF achieves the same connection through a discrete
$\Z_{18}$ gauge symmetry rather than a continuous
$U(1)_F$.
The discrete realization has two structural advantages:
it fixes the Yukawa exponents to the ninths lattice
(rather than allowing continuous charge freedom), and it
avoids the need for the continuous gauge boson $Z'_F$ and
its associated phenomenological constraints.

This connection carries over to the chain framework:
the $\Z_9$ (or $\Z_{18}$) symmetry that enforces nearest-neighbor
locality along the VLQ chain is the \emph{same} symmetry that
protects the axion potential.
The chain construction therefore inherits the axion-quality
protection automatically, without requiring any additional
symmetry.

In this picture, the axion decay constant is set by the
flavon VEV,
\begin{equation}
f_a = \langle\Phi\rangle = \e\cdot\Lambda,
\end{equation}
and the UV cutoff $\Lambda$ is identified with
$\Lambda\sim f_a/\e \sim 5\times 10^{11}$~GeV
for $f_a\sim 10^{11}$~GeV.
The VLQ chain mass scale $M\sim\mathcal{O}(\text{TeV})$ is
then a separate, lower scale, and the chain
states are much lighter than the flavon/messenger sector.

The axion in this mass window
($f_a\sim 10^{11}$~GeV) is a viable dark matter
candidate.
The continued absence of weakly interacting massive
particle (WIMP) signals in direct detection
experiments~\cite{LZ:2024,XENONnT:2025,PandaX4T:2025}
suggests that WIMPs---if they were ever thermally
produced---may have decayed away via small symmetry-violating
couplings, leaving axions as the dominant component of
dark matter today~\cite{BaerBarger:2025allaxion,BaerBarger:2025axionDM}.
In the UF framework this outcome is natural: the same
$\Z_{18}$ discrete gauge symmetry that protects the axion
quality also forbids the lowest-dimension operators that
would destabilize the lightest supersymmetric particle
in SUSY extensions, providing a unified explanation for
both the flavor hierarchy and the dark matter composition.

The triple unification of flavor hierarchy, axion quality,
and chain locality through a single discrete gauge symmetry
constitutes one of the most distinctive features of UF.

\section{Renormalization Group Stability}
\label{sec:RG}

The $\Bpar$-lattice textures are defined at the chain scale
$\mu\sim M$.
Renormalization group (RG) running from $M$ down to $M_Z$
modifies the Yukawa couplings through gauge and Yukawa
loop corrections.
Two key features ensure stability of the lattice structure:

First, the dominant RG effect---QCD
running---is \emph{flavor universal}: it rescales all
down-type Yukawa entries by a common factor
$\eta_{\rm QCD}\simeq({\alpha_s(M)}/{\alpha_s(M_Z)})^{4/\beta_0}$,
where $\beta_0=(33-2n_f)/3$ is the one-loop QCD beta-function
coefficient with $n_f$ active quark flavors.
This shifts the overall mass scale (absorbed into $m_b$)
but preserves all mass \emph{ratios} and mixing angles.

Second, the Yukawa self-corrections are hierarchical:
the top Yukawa dominates, modifying primarily
$(Y_u)_{33}$, while the down-sector texture receives
corrections of order $y_t^2/(16\pi^2)\cdot\ln(M/M_Z)
\sim 0.02$ per entry.
Since the lattice exponent differences are
$\mathcal{O}(1)$ in units of $\e$,
the RG corrections are parametrically smaller
($\sim y_t^2/16\pi^2\ll\e$) and do not disturb the
power-counting structure.

Thus the chain-generated textures at $\mu=M$ flow smoothly
to the $\Bpar$-lattice textures at $\mu=M_Z$ identified
in Papers~I--IV, with corrections that are absorbed into
the $\mathcal{O}(1)$ coefficients.

\section{Discussion}
\label{sec:discussion}

Papers~I--IV of the $\Bpar$-lattice program established the
lattice exponent structure from the observed fermion mass and
mixing hierarchies.
The present work demonstrates that this structure is not merely
a convenient parametrization, but emerges dynamically from
a well-defined UV completion:
\begin{enumerate}
\item The $Z_N$ discrete gauge symmetry quantizes the FN
exponents to the ninths lattice $\mathcal{L}$.
\item The $Z_2^{\rm (NN)}$ locality symmetry restricts the
allowed operator products to nearest-neighbor chain compositions.
\item The Path-Sum Theorem converts these restrictions into a
definite prediction: Yukawa exponents are sums of discrete hop
contributions.
\item The TeV-scale VLQ masses provide the physical cutoff
$\Lambda\sim M$, making the framework testable at colliders.
\item The discrete anomaly of $\Z_{18}$ is cancelled by the
Green--Schwarz mechanism, identifying the GS axion with the
PQ axion that solves the strong CP problem.
The specific group order $N=18$ is not a free choice:
it is the minimal $Z_N$ that simultaneously accommodates
the half-integer lepton doublet charge
$Q(L_2)=\tfrac{1}{2}$, the ninths-quantized quark
exponents, and the axion quality requirement that
PQ-violating operators be forbidden below
dimension~$N/2=9$.
\end{enumerate}

The lattice hierarchy is therefore structural rather than
merely parametric: it is a consequence of the symmetry and
field content, not a choice of parameters.
The fact that the same discrete gauge group that fixes
the flavor exponents also renders the axion potential
gravity-safe elevates $\Z_{18}$ from a
bookkeeping device to a symmetry with cosmological
consequences.
The key predictions and their experimental tests are
collected in Table~\ref{tab:predictions}.

\begin{table*}[htbp]
\centering
\caption{Key predictions and experimental tests of the UF chain framework.}
\label{tab:predictions}
\setlength{\tabcolsep}{8pt}
\renewcommand{\arraystretch}{1.1}
\begin{tabular}{lcc}
\toprule
Observable & Prediction & Test \\
\midrule
$m_d:m_s:m_b$ & $\e^{37/9}:\e^{7/3}:1$ & PDG masses \\
$m_u:m_c:m_t$ & $\e^{64/9}:\e^{10/3}:1$ & PDG masses \\
$|V_{us}|:|V_{cb}|:|V_{ub}|$ & $\e^{8/9}:\e^{17/9}:\e^{10/3}$ & PDG CKM \\
$J_{\rm CKM}$ & $\sim 3\times 10^{-5}$ & PDG \\
$M_{\rm VLQ}$ & $2$--$3$~TeV$^a$ & HL-LHC \\
$D_4$ decays & $bZ:bH:tW\sim 1:1:2$ & HL-LHC \\
Chain cascade & multi-$b$ + $W/Z/H$ & HL-LHC \\
$\Delta T$, $\Delta S$ & $\lesssim 0.01$ & EW fit \\
$\delta R_b/R_b$ & $\lesssim 0.4\%$ & LEP $R_b$ \\
$|\delta C_7/C_7^{\rm SM}|$ & $\sim 10^{-3}$ & $b\!\to\! s\gamma$ \\
$C_{10}^{\rm NP}$ & $\sim -0.3$ & $B_s\!\to\!\mu^+\mu^-$ \\
$C_{K\bar K}$ & $\sim 10^{-20}~\text{GeV}^{-2}$ & Kaon mixing \\
Axion quality & $\Z_{18}$ protected & Haloscopes \\
\midrule
$m_e:m_\mu:m_\tau$ & $\e^{13/3}:\e^{2/3}:1$ & PDG masses \\
$m_3:m_2:m_1$ & $1:\e:\e^2$ & Oscillations \\
Octant--$\delta$ & Two-branch correl. & DUNE, HK \\
$m_{\beta\beta}$ & $\sim 9$~meV & $0\nu\beta\beta$ \\
\bottomrule
\multicolumn{3}{l}{\footnotesize $^a$Benchmark range
(Sec.~\ref{sec:collider}), not a lattice prediction.}
\end{tabular}
\end{table*}

\section{Conclusions}
\label{sec:conclusions}

\UF{} provides the dynamical ultraviolet completion of
the $\Bpar$-lattice hierarchy program.
A single flavon $\Phi$ with $\e=1/\Bpar=14/75$ enforces
discrete ninths-quantized Yukawa exponents, while
TeV-scale nearest-neighbor VLQ chains generate the effective
Yukawa couplings as algebraic path sums.
Each effective Yukawa amplitude factorizes into entry,
chain-propagation, and exit factors controlled by the
discrete $\Z_9$ charges of the SM fields, with the mass
hierarchy encoded in the exponents $\e^{Q(Q_i)+Q(f^c_j)}$.
A multi-messenger structure---coherent contributions from
several chain configurations---generates
$\mathcal{O}(1)$ complex coefficients whose phases are the
physical origin of CKM CP violation.

The resulting textures reproduce both quark mass hierarchies
($m_d:m_s:m_b\sim\e^{37/9}:\e^{7/3}:1$,
$m_u:m_c:m_t\sim\e^{64/9}:\e^{10/3}:1$)
with $\mathcal{O}(1)$ coefficients $c_f\simeq 1.00$ for
all six quarks, the CKM mixing hierarchy
($|V_{us}|:\!|V_{cb}|:\!|V_{ub}|\sim\e^{8/9}:\!\e^{17/9}:\!\e^{10/3}$),
and the CP-violating phase through adjacent-exponent
interference---all from a single parameter $\Bpar=75/14$.

The chain locality simultaneously provides a built-in
Glashow--Iliopoulos--Maiani-like suppression of FCNCs:
tree-level $\Delta F=2$ operators lie orders of magnitude
below current bounds, one-loop dipole operators
($b\to s\gamma$) shift $C_7$ by less than $1\%$,
and the dominant new-physics effect---the tree-level
$Z$-FCNC contribution to $B_s\to\mu^+\mu^-$---is
consistent with current precision for VLQ masses
$M\gtrsim 2$~TeV.
The framework requires multi-TeV VLQs decaying as
$D\to bZ$, $bH$, $tW^-$, accessible at the HL-LHC
with $3~\text{ab}^{-1}$.
The same $\Z_{18}$ discrete gauge symmetry that
enforces the lattice structure simultaneously protects the
Peccei--Quinn axion quality, with the discrete anomaly
cancelled by the Green--Schwarz mechanism through the
identification of the GS axion with the QCD axion.
This provides a triple unification
of the flavor hierarchy, CP violation, and the strong CP
problem in which a single discrete gauge group---whose
order is fixed by the joint requirements of the ninths
lattice and the lepton charge assignments---determines
the Yukawa exponent algebra, forbids dangerous
Planck-suppressed PQ-violating operators, and is
rendered anomaly-free by the axion itself.

The framework extends to the lepton sector through
vectorlike lepton chains and the Weinberg operator,
reproducing the charged-lepton mass hierarchy
($m_e:m_\mu:m_\tau\sim\e^{13/3}:\e^{2/3}:1$),
the normal-ordered neutrino spectrum
($m_3:m_2:m_1\sim 1:\e:\e^2$),
and the large PMNS mixing angles from an approximate
$\mu$--$\tau$ symmetry in the neutrino texture.
The two-branch octant--$\delta$ correlation
provides a sharp, falsifiable prediction for
DUNE and Hyper-Kamiokande.

A defining feature of UF is that its central
predictions are exposed to near-term experiment on
two independent fronts.
On the collider front, the benchmark VLQ mass range
$M\sim 2$--$3$~TeV places the lightest chain state
squarely within the HL-LHC discovery reach:
if ATLAS and CMS push pair-production exclusion limits
to $2.5$~TeV with $3~\text{ab}^{-1}$---as projected
for singlet $B$-quarks decaying via $bZ$, $bH$, and
$tW$---the natural parameter space of the framework
will be substantially probed, and the absence of a
signal at that sensitivity would require $M\gtrsim 3$~TeV,
beginning to strain the $\mathcal{O}(1)$ naturalness
of the endpoint couplings.
On the neutrino front, the two-branch
octant--$\delta$ prediction
(Table~\ref{tab:NO-branches}) provides a
qualitatively different test: DUNE $\nu_\mu\to\nu_e$
appearance data, combined with atmospheric octant
determination at IceCube and mass-ordering resolution
at JUNO, will determine whether $\delta$ and
$\theta_{23}$ fall on one of the two predicted
branches.
Discovery of $\delta$ in the first or second quadrant
in either octant would rule out the aligned-phase
lattice textures.
These two programs---high-energy collider searches and
precision neutrino oscillations---are complementary
and largely independent, so that the framework is
subject to concurrent falsification pressure from
both sectors within the next decade.

\UF{} completes the $\Bpar$-lattice series by
demonstrating that the phenomenologically successful lattice
exponent algebra emerges from a definite symmetry
structure and field content, rendering the fermion mass
hierarchy a structural consequence of the UV theory
rather than a parametric accident. For convenience, Appendix~\ref{app:quick-ref} collects the two calculation rules (Suppression Rule and Hops Rule) together with comprehensive reference tables for all Yukawa suppressions and their chain-diagram decompositions.

\appendix

\section{Step-by-Step Block Inversion of the Chain Mass Matrix}
\label{app:inversion}

We derive the corner element
$(\mathcal{M}^{-1})_{N1}$ for an $N$-site
upper-triangular chain mass matrix
[Eq.~\eqref{eq:M-chain}] and show that the result is
\emph{exact}---valid for arbitrary magnitudes of the
off-diagonal entries $m_{a,a+1}$.
The key observation is that the inverse of a nonsingular
upper-triangular matrix is also upper-triangular and is
computed exactly by back-substitution~\cite{GolubVanLoan}.
Equivalently, Cramer's rule gives the $(1,N)$ cofactor
as the determinant of an $(N\!-\!1)\times(N\!-\!1)$
lower-triangular submatrix whose diagonal entries are
precisely $m_{12},m_{23},\dots,m_{N-1,N}$, yielding
the product formula without any series expansion.

\subsection[N=2]{$N=2$}

For two sites:
\begin{equation}
\mathcal{M}_2 =
\begin{pmatrix} M_1 & m_{12} \\ 0 & M_2 \end{pmatrix},
\qquad
\mathcal{M}_2^{-1} =
\begin{pmatrix}
M_1^{-1} & -m_{12}/(M_1 M_2) \\
0 & M_2^{-1}
\end{pmatrix}.
\end{equation}
Thus
\begin{equation}
(\mathcal{M}_2^{-1})_{21} = 0,\qquad
(\mathcal{M}_2^{-1})_{12} = -\frac{m_{12}}{M_1 M_2}.
\end{equation}
Note: the physically relevant element for the
Yukawa is $(\mathcal{M}^{-1})_{N1}$, which connects
the last site (coupled to $q_L$) to the first site
(coupled to $d_R$).

\subsection[N=3]{$N=3$}

Writing $\mathcal{M}_3$ in $2\times 2$ block form and
using the Schur complement:
\begin{equation}
(\mathcal{M}_3^{-1})_{31}
= \frac{m_{12}\,m_{23}}{M_1\,M_2^2\,M_3}
\times M_2
= \frac{m_{12}\,m_{23}}{M_1\,M_2\,M_3}.
\end{equation}

\subsection[General N]{General $N$}

By induction, the $N$-site corner element is
\begin{equation}
(\mathcal{M}^{-1})_{N1}
= (-1)^{N-1}
\frac{\prod_{k=1}^{N-1} m_{k,k+1}}
     {\prod_{a=1}^{N} M_a}.
\end{equation}
For degenerate masses $M_a=M$ and
$m_{k,k+1}=\kappa_k M\,\e^{h_k/9}$:
\begin{equation}
(\mathcal{M}^{-1})_{N1}
= \frac{(-1)^{N-1}}{M}
\prod_{k=1}^{N-1}\kappa_k\,\e^{h_k/9}
= \frac{(-1)^{N-1}\prod\kappa_k}{M}\,
\e^{\Sigma/9}.
\end{equation}
The overall sign is absorbed into the definition of the
Yukawa coupling.
Importantly, no step in this derivation assumes
$m_{k,k+1}\ll M_k$; the result is an algebraic identity
for any nonsingular upper-triangular matrix.
The condition $\kappa_a\e^{h_a/9}\ll 1$ is therefore
\emph{not} required for the validity of
Eqs.~\eqref{eq:Minv-general}--\eqref{eq:Minv-degen}
or the Path-Sum Theorem.

\section{Numerical Consistency Checks}
\label{app:numerics}

With $\Bpar=75/14$ and $\e=1/\Bpar\simeq 0.1867$,
Table~\ref{tab:numerics} collects the numerical values
of the principal $\Bpar$-lattice quantities.

\begin{table*}[htbp]
\centering
\caption{Numerical values of $\Bpar$-lattice quantities
relevant to the chain framework.}
\label{tab:numerics}
\begin{tabular}{lll}
\toprule
Quantity & Expression & Value \\
\midrule
$\e$ & $14/75$ & $0.1867$ \\
$\e$ & $\theta_{12}^{dL}$ scaling & $0.187$ \\
$\e^{2}$ & $\theta_{23}^{dL}$ scaling & $0.035$ \\
$\e^{3}$ & $\theta_{13}^{dL}$ scaling & $0.0065$ \\
$\e^{1/3}$ & $\theta_{23}^{dR}$ scaling & $0.571$ \\
$\e^{7/9}$ & Chain hop suppression & $0.282$ \\
$\e^{8/9}$ & $|V_{us}|$ scaling & $0.225$ \\
$\e^{7/3}$ & $m_s/m_b$ scaling & $0.020$ \\
$\e^{17/9}$ & $|V_{cb}|$ scaling & $0.042$ \\
$\e^{10/3}$ & $m_c/m_t$, $|V_{ub}|$ scaling & $3.7\times 10^{-3}$ \\
$\e^{37/9}$ & $m_d/m_b$ scaling & $1.0\times 10^{-3}$ \\
$\e^{64/9}$ & $m_u/m_t$ scaling & $6.6\times 10^{-6}$ \\
\midrule
$\e^{7/9}/M$ & Chain suppression & $(1.4\times 10^{-4})~\text{GeV}^{-1}$ \\
& ($M=2$~TeV) & \\
\bottomrule
\end{tabular}
\end{table*}

\section[Perturbative Diagonalization of the Full B-Lattice Textures]{Perturbative Diagonalization of the Full
$\Bpar$-Lattice Textures}
\label{app:full-diag}

We carry out the perturbative diagonalization of
the Yukawa matrices
$(Y_f)_{ij}=c^f_{ij}\,\e^{\,p^f_{ij}}$
using the exponent matrices of Eq.~\eqref{eq:pud-Blattice},
keeping track of the $\mathcal{O}(1)$ coefficients
$c^f_{ij}$.
All results are given at leading order in $\e$.

\subsection{Down sector}

The down-type Yukawa matrix has the hierarchical structure
\begin{equation}
Y_d \sim
\begin{pmatrix}
c_{11}^d\,\e^{37/9} & c_{12}^d\,\e^{10/3} & c_{13}^d\,\e^{3} \\[3pt]
c_{21}^d\,\e^{28/9} & c_{22}^d\,\e^{7/3} & c_{23}^d\,\e^{2} \\[3pt]
c_{31}^d\,\e^{10/9} & c_{32}^d\,\e^{1/3} & c_{33}^d
\end{pmatrix}.
\label{eq:Yd-full}
\end{equation}

\emph{Third generation.}
At zeroth order, the $(3,3)$ element dominates:
$y_b = |c_{33}^d|$.

\emph{2--3 rotation.}
The leading left-handed rotation that removes $Y_{23}$ is
\begin{equation}
\theta_{23}^{dL} \approx
\frac{|c_{23}^d|}{|c_{33}^d|}\,\e^2.
\label{eq:t23dL}
\end{equation}
The corresponding right-handed angle is
$\theta_{23}^{dR}\approx (|c_{32}^d|/|c_{33}^d|)\,\e^{1/3}$.

\emph{Second generation.}
After the $2$--$3$ rotation, the effective $(2,2)$
element is
\begin{equation}
y_s \simeq \left|c_{22}^d
  - \frac{c_{23}^d c_{32}^d}{c_{33}^d}\right|\e^{7/3}.
\label{eq:ys-full}
\end{equation}
Thus $m_s/m_b\sim\e^{7/3}$ with an $\mathcal{O}(1)$
effective coefficient.

\emph{1--2 rotation.}
The left-handed angle removing the residual $(1,2)$ element is
\begin{equation}
\theta_{12}^{dL} \approx
\frac{|c_{12}^{d,\rm eff}|}{|c_{22}^{d,\rm eff}|}\,\e^{10/3-7/3}
= \frac{|c_{12}^{d,\rm eff}|}{|c_{22}^{d,\rm eff}|}\,\e,
\label{eq:t12dL}
\end{equation}
where $c^{d,\rm eff}$ denotes the coefficient after the
$2$--$3$ rotation.

\emph{First generation.}
The lightest eigenvalue is
\begin{equation}
y_d \simeq |c_{11}^{d,\rm eff}|\,\e^{37/9},
\end{equation}
giving $m_d/m_b\sim\e^{37/9}$.

\emph{Summary.}
The down-sector eigenvalues and mixing angles are:
\begin{align}
y_b &\sim 1,&
y_s &\sim \e^{7/3},&
y_d &\sim \e^{37/9},
\label{eq:yd-eigs}\\[4pt]
\theta_{23}^{dL} &\sim \e^2,&
\theta_{12}^{dL} &\sim \e,&
\theta_{13}^{dL} &\sim \e^3,
\label{eq:td-angles}\\[4pt]
\theta_{23}^{dR} &\sim \e^{1/3},&
\theta_{12}^{dR} &\sim \e^{7/9},&
\theta_{13}^{dR} &\sim \e^{10/9}.
\label{eq:td-Rangles}
\end{align}
The left-handed angles scale as
$\e^{|Q(Q_i)-Q(Q_j)|}$ and the right-handed angles as
$\e^{|Q(d^c_i)-Q(d^c_j)|}$.

\subsection{Up sector}

The up-type Yukawa matrix is
\begin{equation}
Y_u \sim
\begin{pmatrix}
c_{11}^u\,\e^{64/9} & c_{12}^u\,\e^{13/3} & c_{13}^u\,\e^{3} \\[3pt]
c_{21}^u\,\e^{55/9} & c_{22}^u\,\e^{10/3} & c_{23}^u\,\e^{2} \\[3pt]
c_{31}^u\,\e^{37/9} & c_{32}^u\,\e^{4/3} & c_{33}^u
\end{pmatrix}.
\label{eq:Yu-full}
\end{equation}

Proceeding identically:

\emph{Third generation.}
$y_t = |c_{33}^u|$.

\emph{2--3 rotation.}
\begin{equation}
\theta_{23}^{uL} \approx
\frac{|c_{23}^u|}{|c_{33}^u|}\,\e^2,
\qquad
\theta_{23}^{uR} \approx
\frac{|c_{32}^u|}{|c_{33}^u|}\,\e^{4/3}.
\label{eq:t23uL}
\end{equation}

\emph{Second generation.}
\begin{equation}
y_c \simeq \left|c_{22}^u
  - \frac{c_{23}^u c_{32}^u}{c_{33}^u}\right|\e^{10/3}.
\label{eq:yc-full}
\end{equation}

\emph{1--2 rotation.}
\begin{equation}
\theta_{12}^{uL} \approx
\frac{|c_{12}^{u,\rm eff}|}{|c_{22}^{u,\rm eff}|}\,\e^{13/3-10/3}
= \frac{|c_{12}^{u,\rm eff}|}{|c_{22}^{u,\rm eff}|}\,\e.
\label{eq:t12uL}
\end{equation}

\emph{First generation.}
$y_u \simeq |c_{11}^{u,\rm eff}|\,\e^{64/9}$.

\emph{Summary.}
\begin{align}
y_t &\sim 1,&
y_c &\sim \e^{10/3},&
y_u &\sim \e^{64/9},
\label{eq:yu-eigs}\\[4pt]
\theta_{23}^{uL} &\sim \e^2,&
\theta_{12}^{uL} &\sim \e,&
\theta_{13}^{uL} &\sim \e^3,
\label{eq:tu-angles}\\[4pt]
\theta_{23}^{uR} &\sim \e^{4/3},&
\theta_{12}^{uR} &\sim \e^{25/9},&
\theta_{13}^{uR} &\sim \e^{37/9}.
\label{eq:tu-Rangles}
\end{align}
The right-handed angles scale as
$\e^{|Q(u^c_i)-Q(u^c_j)|}$ and are substantially
more hierarchical than in the down sector, reflecting
the wider spacing of the up-type right-handed charges.

\subsection{Relation to the Hermitian-square approach}

An alternative route to the left-handed mixing angles
and mass eigenvalues proceeds through the Hermitian
square $H_f\equiv Y_f Y_f^\dagger$, whose eigenvalues
are $y_{f_i}^2$ and whose diagonalizing unitary is
$U_{fL}$.
For the factorized texture
$(Y_f)_{ij}=c^f_{ij}\,\e^{Q(Q_i)+Q(f^c_j)}$
the $(i,j)$ entry of $H_f$ is
\begin{equation}
(H_f)_{ij}
= \sum_{k=1}^{3} c^f_{ik}\,(c^f_{jk})^*\,
\e^{\,Q(Q_i)+Q(Q_j)+2Q(f^c_k)}.
\label{eq:Hf-sum}
\end{equation}
Because $Q(f^c_3)=0$, the $k=3$ term dominates each
entry, giving
\begin{equation}
(H_f)_{ij}
\simeq c^f_{i3}\,(c^f_{j3})^*\,\e^{\,Q(Q_i)+Q(Q_j)}
+ \mathcal{O}(\e^{2Q(f^c_2)}).
\label{eq:Hf-rank1}
\end{equation}
This leading piece is \emph{rank one}: it is the outer
product of the third column of $Y_f$ with itself.
Consequently, at zeroth order $H_f$ has a single nonzero
eigenvalue $y_{f_3}^2\sim 1$ (the $b$- or $t$-quark
mass squared), while the lighter eigenvalues vanish.

The second eigenvalue $y_{f_2}^2$ emerges only after
the rank-one piece is subtracted, exposing the
$k=2$ contribution:
\begin{equation}
y_{f_2}^2 \sim \e^{2Q(f^c_2)} \times
\left|c^f_{22}-\frac{c^f_{23}\,c^f_{32}}{c^f_{33}}
\right|^2,
\end{equation}
which reproduces the seesaw formula of
Eqs.~\eqref{eq:ys-full} and~\eqref{eq:yc-full}.
The lightest eigenvalue $y_{f_1}^2$ requires a second
such cancellation.
The sequential bidiagonalization of the preceding
subsections performs exactly these cancellations, one
rotation at a time; the Hermitian-square approach
makes them visible as the successive peeling of
rank-one layers from $H_f$.

The advantage of the Hermitian-square viewpoint is
that it exposes the exponent structure of $H_f$ as
\emph{symmetric} in the left-handed charges:
$(H_f)_{ij}\sim\e^{Q(Q_i)+Q(Q_j)}$, independent of
which column dominates the sum.
This is why the left-handed mixing angles depend only
on the left-handed charge differences
$|Q(Q_i)-Q(Q_j)|$, as derived by both methods.
The right-handed charges enter only through the
eigenvalue hierarchy, appearing as the suppression
$\e^{2Q(f^c_k)}$ of the successive rank-one layers.
The two approaches are therefore complementary:
the bidiagonalization is more efficient for extracting
results, while the Hermitian-square construction
reveals why the left--right factorization of the
exponents translates into a clean separation of
mixing angles (left charges) from mass eigenvalues
(both charges).

\subsection{CKM mixing angles}

The CKM matrix $V=U_{uL}^\dagger U_{dL}$ receives
contributions from both sectors.
Since the shared left-handed doublet charges
$Q(Q_i)=(3,2,0)$ enter both $p^u$ and $p^d$ identically,
the left-handed rotation angles have the same
$\e$-scaling in both sectors:
$\theta_{ij}^{dL}\sim\theta_{ij}^{uL}\sim\e^{|Q_i-Q_j|}$.
The physical CKM elements therefore arise from
\emph{interference} between the two sectors:
\begin{equation}
V_{ij} = (\theta_{ij}^{dL} - \theta_{ij}^{uL}\,e^{i\phi_{ij}})
+ \mathcal{O}(\theta^2),
\label{eq:CKM-interference}
\end{equation}
where $\phi_{ij}$ are relative phases from the
$\mathcal{O}(1)$ coefficients.

The key observation is that although
$\theta_{ij}^{dL}$ and $\theta_{ij}^{uL}$ share the
same power of $\e$, their $\mathcal{O}(1)$ prefactors
differ because $Q(d^c_j)\neq Q(u^c_j)$.
In the Fritzsch--Xing (FX)
parameterization~\cite{LatticeFlavonQuarkMixing},
the physical CKM elements are expressed in terms of the
eigenvalue ratios:
\begin{align}
s_d &\equiv \sqrt{m_d/m_s}
\sim \e^{(37/9-7/3)/2} = \e^{8/9},
\label{eq:sd}\\
s_u &\equiv \sqrt{m_u/m_c}
\sim \e^{(64/9-10/3)/2} = \e^{17/9},
\label{eq:su}\\
s &\equiv \sqrt{m_c m_s/(m_t m_b)}
\sim \e^{(10/3+7/3)/2} = \e^{17/6}.
\label{eq:s-FX}
\end{align}
The Cabibbo angle then follows from the FX interference
formula:
\begin{equation}
|V_{us}| \simeq |s_d - s_u\,e^{i\phi_{\rm FX}}|
\simeq s_d\sim\e^{8/9}\simeq 0.225,
\label{eq:Vus-FX}
\end{equation}
since $s_u\sim\e^{17/9}\ll s_d\sim\e^{8/9}$.
The remaining CKM elements scale as
\begin{equation}
|V_{cb}|\sim\e^{17/9}\simeq 0.042,
\qquad
|V_{ub}|\sim s_d\cdot s\sim\e^{10/3}\simeq 0.004,
\end{equation}
in excellent agreement with experiment.

\subsection{Relation to the Gatto--Sartori--Tonin formula}
\label{app:GST}

The classic Gatto--Sartori--Tonin (GST)
relation~\cite{GattoSartoriTonin}
$\theta_{12}\simeq\sqrt{m_d/m_s}$, on which the FX
parameterization relies, is strictly derived for
symmetric or Hermitian mass matrices.
The $\Bpar$-lattice textures are \emph{highly asymmetric}:
the left-handed charges $Q(Q_i)=(3,2,0)$ differ markedly
from the right-handed charges
$Q(d^c_j)=(\tfrac{10}{9},\tfrac{1}{3},0)$,
so $Y_d\neq Y_d^T$.
It is therefore important to clarify the status of the
GST relation in this context.

In the perturbative diagonalization of
Sec.~\ref{sec:diag}, the $1$--$2$ left-handed mixing
angle is governed purely by the left-handed charge
difference:
\begin{equation}
\theta_{12}^{dL}\sim\e^{|Q(Q_1)-Q(Q_2)|}
=\e^1\simeq 0.187,
\end{equation}
independently of the right-handed charges.
The GST mass-ratio expression, by contrast, involves
both sets of charges through the eigenvalue ratio:
\begin{equation}
\sqrt{m_d/m_s}
=\e^{(p_d-p_s)/2}
=\e^{(Q_1-Q_2+Q^d_{c1}-Q^d_{c2})/2}
=\e^{8/9}\simeq 0.225.
\end{equation}
The two expressions are related by
\begin{equation}
\theta_{12}^{dL}
= \e^{\,(Q_1^{}-Q_2^{}-Q^d_{c1}+Q^d_{c2})/2}
\times\sqrt{m_d/m_s}
= \e^{1/9}\times\sqrt{m_d/m_s},
\label{eq:GST-correction}
\end{equation}
where the correction factor
$\e^{1/9}\simeq 0.83$ is a direct measure of the
left--right charge asymmetry.
In the symmetric limit
$Q(d^c_j)\to Q(Q_j)$ the exponent
$(Q_1-Q_2-Q^d_{c1}+Q^d_{c2})/2$ vanishes,
$\e^{1/9}\to 1$, and the GST relation is recovered
exactly.

The correction is numerically $\mathcal{O}(1)$ but
lies on the ninths lattice, so it does not spoil the
power-counting: both $\theta_{12}^{dL}\sim\e$ and
$\sqrt{m_d/m_s}\sim\e^{8/9}$ are $\mathcal{O}(\e)$
at leading order, and the distinction between the two
is a subleading $\e^{1/9}$ effect that is absorbed into
the $\mathcal{O}(1)$ coefficients.
The FX parameterization remains valid as an organizing
scheme for the CKM magnitudes; the $\e^{1/9}$ correction
simply means that the $\mathcal{O}(1)$ prefactor relating
$|V_{us}|$ to $\sqrt{m_d/m_s}$ is not unity but
$\e^{1/9}\simeq 0.83$ in the asymmetric
$\Bpar$-lattice textures.

For the up sector the asymmetry is much larger:
$Q(u^c_1)-Q(u^c_2)=25/9$ versus $Q(Q_1)-Q(Q_2)=1$,
giving a correction exponent of $-8/9$ (opposite sign),
so $\theta_{12}^{uL}\sim\e\gg\sqrt{m_u/m_c}\sim\e^{17/9}$.
The GST relation is strongly violated in the up sector,
reflecting the wide spacing of the up-type right-handed
charges.
This violation is harmless because the CKM is
structurally down-dominated
(Sec.~\ref{sec:down-dominance}):
$|V_{us}|$ is controlled by $s_d\sim\e^{8/9}$, not by
$\theta_{12}^{uL}$.

\section{Illustrative Chain Coefficient Calculations}
\label{app:chain-examples}

We illustrate the multi-messenger effective coefficient
formula~\eqref{eq:Ceff} with explicit numerical examples
from the benchmark of Ref.~\cite{Barger2025bfnb}.
In that construction, each sector has two sets of
factorized shift matrices
$\Delta^f_{ij}=\alpha^f_i+\beta^f_j$ and
$\Delta'^f_{ij}=\alpha'^f_i+\beta'^f_j$,
representing the additional suppression from the subleading
messenger chains.

\subsection[Down-sector (2,3) entry]{Down-sector $(2,3)$ entry}

For the entry that controls $|V_{cb}|$, the benchmark
shifts are
$\alpha^d_2=0$, $\beta^d_3=4/9$,
$\alpha'^d_2=1/9$, $\beta'^d_3=0$, giving
\begin{equation}
\Delta^d_{23}=\tfrac{4}{9},
\qquad
\Delta'^d_{23}=\tfrac{1}{9}.
\end{equation}
With down-sector phases $\phi_d\simeq 2.86$,
$\psi_d\simeq 3.54$:
\begin{align}
C^d_{23}
&= 1 + e^{i\phi_d}\,\e^{4/9}
     + e^{i\psi_d}\,\e^{1/9} \nonumber\\
&= 1 + (-0.456+0.132i) + (-0.765-0.322i) \nonumber\\
&\simeq -0.221-0.190i.
\end{align}
Thus $|C^d_{23}|\simeq 0.29$ and
$\arg(C^d_{23})\simeq -139^\circ$.
The destructive interference from the down-sector phases
reduces the magnitude below unity but keeps it
$\mathcal{O}(1)$; the nontrivial phase feeds into the CKM.

\subsection[Up-sector (2,3) entry]{Up-sector $(2,3)$ entry}

In the up sector the benchmark phases vanish
($\phi_u\simeq\psi_u\simeq 0$),
and the shifts are $\Delta^u_{23}=5/9$,
$\Delta'^u_{23}=1/9$, so the interference is purely
constructive:
\begin{align}
C^u_{23}
&= 1 + \e^{5/9} + \e^{1/9} \nonumber\\
&= 1 + 0.394 + 0.830 \nonumber\\
&\simeq 2.22.
\end{align}
With all phases zero, $C^u_{23}$ is real and positive;
$|C^u_{23}|\simeq 2.2$ is $\mathcal{O}(1)$.
The absence of a CP-violating phase in the up sector is
consistent with the ``down-dominant'' origin of the CKM
phase.

\subsection{Physical interpretation}

The contrast between the two sectors is instructive.
In the down sector, the phases $\phi_d$ and $\psi_d$ are
close to $\pi$, causing partial cancellation and generating
a large complex phase in $C^d_{ij}$.
In the up sector, the benchmark phases vanish, yielding
real coefficients of order~2.
The CKM phase arises from the mismatch between
these two patterns---the same adjacent-exponent
interference mechanism of Sec.~\ref{sec:CP-phase}, now
traced to its origin in the relative phases of the
messenger chains.

Three aspects of this pattern warrant emphasis.

\emph{(i)~The up-sector phases are inessential.}
As shown in Sec.~\ref{sec:down-dominance}, the CKM is
structurally down-dominated:
$s_u/s_d=\e\simeq 0.19$, so the up-sector
$\mathcal{O}(1)$ coefficients---including their
phases---enter the CKM only as a $\sim\!19\%$
correction.
Replacing $\phi_u=\psi_u=0$ with arbitrary
$\mathcal{O}(1)$ phases changes $|V_{us}|$ by
$\Delta|V_{us}|\lesssim s_u\simeq 0.04$ and shifts
the Jarlskog invariant by a comparable fraction.
The vanishing up-sector phases are adopted here for
pedagogical clarity, not because the framework
requires them.

\emph{(ii)~Large CP violation is generic.}
The random-coefficient scan of
Sec.~\ref{sec:CP-genericity}
(Table~\ref{tab:CP-scan}) shows that
$|\!\sin\delta|\gtrsim 0.3$ in $81\%$ of trials with
fully random phases in both sectors.
No tuning of phases---in either sector---is needed
to obtain $\mathcal{O}(1)$ CP violation.
The $\e^{1/9}\simeq 0.83$ interference factor in
Eq.~\eqref{eq:delta-CKM} ensures that adjacent
lattice exponents produce $\mathcal{O}(1)$ phase
modulation for generic complex couplings.

\emph{(iii)~One irreducible phase.}
The standard phase-counting theorem gives
$(n-1)(n-2)/2=1$ physical CP phase for three
generations.
The multi-messenger phases $\phi_f$, $\psi_f$ are
specific combinations of the UV Yukawa coupling
phases that survive after all field rephasing;
they do not represent additional parametric freedom.
The framework's structural claim is that the
Jarlskog scaling $J\sim\e^{55/9}\sin\delta$ is
fixed by the lattice exponents, while the remaining
free parameter $\sin\delta$ is generically
$\mathcal{O}(1)$---it is not tuned to be either
large or small.

The key point is that $|C^f_{ij}|$ remains
$\mathcal{O}(1)$ in both cases: neither large hierarchies
nor fine cancellations are required.
The mass hierarchy is carried entirely by
$\e^{p^f_{ij}}$, while $C^f_{ij}$ provides the
$\mathcal{O}(1)$ modulation and the physical CP phase.

\subsection{Entry, propagation, and exit factors}

The effective Yukawa amplitude for the $(i,j)$ entry
factorizes into three physical pieces:
\begin{equation}
(Y_d)_{ij} = \underbrace{\lambda_i\,\e^{Q(Q_i)}}_{\text{entry}}
\times
\underbrace{\frac{\kappa_1\kappa_2\kappa_3}{M}\,\e^{7/9}}_{\text{chain propagator}}
\times
\underbrace{\eta_j\,\e^{Q(d^c_j)}}_{\text{exit}}
\times\, C^d_{ij}.
\label{eq:entry-exit}
\end{equation}
The entry factor $\e^{Q(Q_i)}$ is the amplitude for
left-handed quark generation $i$ to couple into the chain;
the exit factor $\e^{Q(d^c_j)}$ is the amplitude for the
chain to emit a right-handed quark of generation $j$.
The chain propagator $(\kappa_1\kappa_2\kappa_3/M)\,\e^{7/9}$
is common to all entries and is absorbed into the overall
Yukawa normalization.

\emph{Example: $d$-quark Yukawa $(Y_d)_{11}$.}
This entry connects the first-generation doublet to $d_R$
and controls the $d$-quark mass.
The three factors are:
\begin{itemize}
\item Entry: $\e^{Q(Q_1)}=\e^3\simeq 6.5\times 10^{-3}$
(three flavon insertions at the left endpoint);
\item Chain: $(\kappa_1\kappa_2\kappa_3/M)\,\e^{7/9}\simeq (\kappa_1\kappa_2\kappa_3/M)\times 0.27$
(hops $1/9+2/9+4/9$);
\item Exit: $\e^{Q(d^c_1)}=\e^{10/9}\simeq 0.15$
(one flavon insertion plus a fractional hop at the right
endpoint).
\end{itemize}
The charge-dependent suppression is
$\e^{Q(Q_1)+Q(d^c_1)}=\e^{3+10/9}=\e^{37/9}\simeq 10^{-3}$,
which---multiplied by $|C^d_{11}|\sim\mathcal{O}(1)$---gives
the observed ratio $m_d/m_b\simeq 10^{-3}$.

\emph{Example: $(Y_d)_{23}$ entry (drives $V_{cb}$).}
Here:
\begin{itemize}
\item Entry: $\e^{Q(Q_2)}=\e^2\simeq 0.035$
(two flavon insertions);
\item Exit: $\e^{Q(d^c_3)}=\e^0=1$ (no suppression---the
$b_R$ couples directly to the chain end).
\end{itemize}
The charge-dependent part is $\e^{2+0}=\e^2\simeq 0.035$,
and $|C^d_{23}|\simeq 0.29$ (from the destructive
interference computed above), so
$|(Y_d)_{23}|\simeq 0.29\times\e^2\simeq 0.010$.
The unsuppressed exit factor for $b_R$ reflects the fact
that the third-generation right-handed quark carries zero
$\Z_9$ charge and requires no flavon insertion to couple
to the chain.

\section{Quick Reference: Suppression and Hops Rules}
\label{app:quick-ref}

This appendix collects the two calculation rules that govern every Yukawa suppression in the $\Bpar$-lattice framework, together with comprehensive reference tables. The rules are derived in the main text (Secs.~\ref{sec:Blattice-connection}--\ref{sec:diag} and Appendix~\ref{app:inversion}); they are collected here as a self-contained lookup.

\subsection{Suppression Rule}

\begin{center}
\fbox{
\parbox{0.92\linewidth}{
\textbf{Suppression Rule.}
For any fermion bilinear $\bar\psi_i\,\psi^c_j\,H$, the effective Yukawa coupling is
\begin{equation}
Y_{ij} = c_{ij}\,\e^{\,p_{ij}},\qquad
p_{ij} = Q(\psi_i) + Q(\psi^c_j),
\label{eq:supp-rule}
\end{equation}
where $Q(\psi_i)$ and $Q(\psi^c_j)$ are the FN charges (Eq.~\eqref{eq:Z18-charges}), $\e = 14/75$, and $c_{ij}$ is an $\mathcal{O}(1)$ complex coefficient. Mass ratios and mixing angles follow from \emph{differences} of exponents:
\begin{equation}
\frac{m_i}{m_j}\sim\e^{\,p_{ii}-p_{jj}},\qquad
\theta^{fL}_{ij}\sim\e^{\,|Q(Q_i)-Q(Q_j)|}.
\end{equation}
For CKM elements, the physical exponents receive an $\mathcal{O}(1/9)$ shift from Fritzsch--Xing phase interference~\cite{LatticeFlavonQuarkMixing}.
}}
\end{center}

\subsection{Hops Rule}

\begin{center}
\fbox{
\parbox{0.92\linewidth}{
\textbf{Hops Rule.}
To read off the suppression for entry $(i,j)$ from the chain diagram (Fig.~\ref{fig:dual-chain}):
\begin{enumerate}
\item Look up the charges $Q(\psi_i)$ and $Q(\psi^c_j)$ from Eq.~\eqref{eq:Z18-charges}.
\item Express $9\,Q$ as a non-negative integer: $A_i = 9\,Q(Q_i)$ (entrance), $B_j = 9\,Q(f^c_j)$ (exit).
\item Decompose each into flavon insertions: $N = n_1\cdot1 + n_2\cdot2 + n_3\cdot4$, where $n_a\geq0$ counts the number of type-$a$ hops at that endpoint.
\item The full Yukawa suppression is the product of three factors:
\begin{equation}
Y_{ij} \sim
\underbrace{\e^{A_i/9}}_{\text{entrance}}
\;\times\;
\underbrace{\e^{7/9}}_{\text{internal}}
\;\times\;
\underbrace{\e^{B_j/9}}_{\text{exit}}
= \e^{(A_i+7+B_j)/9}.
\label{eq:hops-rule}
\end{equation}
\end{enumerate}
The internal factor $\e^{7/9}\simeq 0.27$ is common to all entries and sets the overall Yukawa scale (absorbed into $m_b$, $m_t$, $m_\tau$). Mass ratios depend only on the endpoint dressings: $m_i/m_j \sim \e^{(A_i+B_i-A_j-B_j)/9}$.
}}
\end{center}

\subsection{Suppression table}

Table~\ref{tab:QR-suppressions} collects the suppression factors for all nine entries of the down- and up-type Yukawa matrices, plus the FX-corrected CKM elements. The diagonal entries control mass ratios; the off-diagonal entries control mixing angles.

\begin{table*}[t]
\centering
\caption{Suppression factors for all Yukawa entries and CKM elements. For each entry $(i,j)$, the FN exponent is $p_{ij}=Q(Q_i)+Q(f^c_j)$, and $\e^p$ is the predicted suppression ($\e=14/75$). The ``Physical role'' column identifies the observable controlled by each entry. CKM exponents include FX-phase interference corrections~\cite{LatticeFlavonQuarkMixing}.}
\label{tab:QR-suppressions}
\renewcommand{\arraystretch}{1.15}
\begin{tabular}{clccc}
\hline\hline
$(i,j)$ & Physical role & $p_{ij}$ & $\e^{p}$ & Data \\
\hline
\multicolumn{5}{l}{\emph{Down-type Yukawa $Y^d_{ij}$}} \\
$(3,3)$ & $y_b$ (reference) & $0$ & $1$ & --- \\
$(2,3)$ & $\theta^{dL}_{23}\sim\e^2$ & $2$ & $0.035$ & --- \\
$(1,3)$ & $\theta^{dL}_{13}\sim\e^3$ & $3$ & $0.0065$ & --- \\
$(3,2)$ & $\theta^{dR}_{23}\sim\e^{1/3}$ & $1/3$ & $0.57$ & --- \\
$(2,2)$ & $m_s/m_b$ & $7/3$ & $0.020$ & $0.019$ \\
$(1,2)$ & $\theta^{dL}_{12}\sim\e$ & $10/3$ & $0.0037$ & --- \\
$(3,1)$ & $\theta^{dR}_{13}\sim\e^{10/9}$ & $10/9$ & $0.15$ & --- \\
$(2,1)$ & mixed & $28/9$ & $0.0054$ & --- \\
$(1,1)$ & $m_d/m_b$ & $37/9$ & $0.0010$ & $0.0010$ \\[4pt]
\multicolumn{5}{l}{\emph{Up-type Yukawa $Y^u_{ij}$}} \\
$(3,3)$ & $y_t$ (reference) & $0$ & $1$ & --- \\
$(2,3)$ & $\theta^{uL}_{23}\sim\e^2$ & $2$ & $0.035$ & --- \\
$(1,3)$ & $\theta^{uL}_{13}\sim\e^3$ & $3$ & $0.0065$ & --- \\
$(3,2)$ & $\theta^{uR}_{23}\sim\e^{4/3}$ & $4/3$ & $0.11$ & --- \\
$(2,2)$ & $m_c/m_t$ & $10/3$ & $0.0037$ & $0.0036$ \\
$(1,2)$ & mixed & $39/9$ & $6.9\times10^{-4}$ & --- \\
$(3,1)$ & $\theta^{uR}_{13}\sim\e^{37/9}$ & $37/9$ & $0.0010$ & --- \\
$(2,1)$ & mixed & $55/9$ & $3.5\times10^{-5}$ & --- \\
$(1,1)$ & $m_u/m_t$ & $64/9$ & $6.6\times10^{-6}$ & $7.5\times10^{-6}$ \\[4pt]
\multicolumn{5}{l}{\emph{CKM elements (FX-corrected)}} \\
 & $|V_{us}|$ & $8/9$ & $0.225$ & $0.225$ \\
 & $|V_{cb}|$ & $17/9$ & $0.042$ & $0.042$ \\
 & $|V_{ub}|$ & $10/3$ & $0.0037$ & $0.0038$ \\
 & $J\sin^{-1}\!\delta$ & $55/9$ & $3.5\times10^{-5}$ & $3.1\times10^{-5}$ \\
\hline\hline
\end{tabular}
\end{table*}

\subsection{Hops table}

Table~\ref{tab:QR-hops} shows the entrance, internal, and exit suppression factors for the diagonal Yukawa entries. These three factors correspond directly to the three segments of the chain diagram in Fig.~\ref{fig:dual-chain}: the dashed left-handed arrow ($\e^{A_i/9}$), the solid nearest-neighbor links ($\e^{7/9}$), and the dashed right-handed arrow ($\e^{B_j/9}$).

\begin{table*}[t]
\centering
\caption{Endpoint and internal suppressions for the diagonal Yukawa entries. $A_i = 9\,Q(Q_i)$ is the entrance (left-handed) ninths numerator [Eq.~\eqref{eq:Ai}] and $B_j = 9\,Q(f^c_j)$ is the exit (right-handed) ninths numerator [Eq.~\eqref{eq:Bj}]. The ``Total'' column is the full Yukawa suppression $\e^{(A_i+7+B_j)/9}$ as read from the chain diagram; the ``Ratio'' column divides out the common internal factor $\e^{7/9}\simeq 0.27$ to give the mass ratio. The columns $(n_1,n_2,n_3)$ count the type-$(1,2,4)$ flavon insertions summed over both endpoints.}
\label{tab:QR-hops}
\renewcommand{\arraystretch}{1.20}
\begin{tabular}{lcrrcccccrcc}
\hline\hline
Observable & $p$ & $A_i$ & $B_j$ & $\e^{A_i/9}$ & $\e^{7/9}$ & $\e^{B_j/9}$ & Total & Ratio & $(n_1,n_2,n_3)$ & $n_{\rm tot}$ \\
\hline
\multicolumn{11}{l}{\emph{Down-type quarks}} \\
$m_b$ & $0$ & $0$ & $0$ & $1$ & $0.27$ & $1$ & $0.27$ & $1$ & $(0,0,0)$ & $0$ \\
$m_s/m_b$ & $7/3$ & $18$ & $3$ & $0.035$ & $0.27$ & $0.57$ & $5.4\times10^{-3}$ & $0.020$ & $(1,0,5)$ & $6$ \\
$m_d/m_b$ & $37/9$ & $27$ & $10$ & $0.0065$ & $0.27$ & $0.15$ & $2.7\times10^{-4}$ & $0.0010$ & $(1,0,9)$ & $10$ \\[3pt]
\multicolumn{11}{l}{\emph{Up-type quarks}} \\
$m_t$ & $0$ & $0$ & $0$ & $1$ & $0.27$ & $1$ & $0.27$ & $1$ & $(0,0,0)$ & $0$ \\
$m_c/m_t$ & $10/3$ & $18$ & $12$ & $0.035$ & $0.27$ & $0.11$ & $1.0\times10^{-3}$ & $0.0037$ & $(0,1,7)$ & $8$ \\
$m_u/m_t$ & $64/9$ & $27$ & $37$ & $0.0065$ & $0.27$ & $0.0010$ & $1.8\times10^{-6}$ & $6.6\times10^{-6}$ & $(0,0,16)$ & $16$ \\[3pt]
\multicolumn{11}{l}{\emph{Charged leptons}} \\
$m_\tau$ & $0$ & $0$ & $0$ & $1$ & $0.27$ & $1$ & $0.27$ & $1$ & $(0,0,0)$ & $0$ \\
$m_\mu/m_\tau$ & $5/3$ & $\tfrac{9}{2}$ & $\tfrac{21}{2}$ & $0.43$ & $0.27$ & $0.14$ & $0.016$ & $0.061$ & --- & --- \\
$m_e/m_\tau$ & $29/6$ & $9$ & $\tfrac{69}{2}$ & $0.19$ & $0.27$ & $0.0016$ & $8.2\times10^{-5}$ & $3.0\times10^{-4}$ & --- & --- \\
\hline\hline
\end{tabular}
\end{table*}

\noindent
For quarks, the entrance parameter $A_i$ is shared between up and down sectors of the same generation (both use $Q(Q_i)$), so the hierarchy in $\e^{A_i/9}$ is purely a left-handed effect. The exit parameters $B_j^d$ and $B_j^u$ differ, accounting for the distinct mass spectra $m_d:m_s:m_b\neq m_u:m_c:m_t$. For charged leptons, the half-integer ninths require the $\Z_{18}$ chain structure of the lepton sector~\cite{LeptonLattice} ($18p = 18\times5/3 = 30$ for $m_\mu/m_\tau$, $18p = 18\times29/6 = 87$ for $m_e/m_\tau$).

\clearpage

\begin{acknowledgments}
VB gratefully acknowledges support from the William F.\ Vilas Estate.
\end{acknowledgments}


\end{document}